%% file: main.tex
\documentclass[usenatbib]{mnras}
\usepackage[utf8]{inputenc}
\usepackage[english]{babel}
\usepackage[utf8]{inputenc}
\usepackage[T1]{fontenc}
\usepackage{ae,aecompl}

\usepackage{amsmath}
\usepackage{dsfont}
\usepackage{amssymb}

\usepackage{gensymb}

\usepackage{graphicx}
\usepackage{natbib}
\usepackage{url}

\usepackage{tikz}
\usetikzlibrary{arrows,shapes}

\usepackage{rotating}
\usepackage{xcolor}
\usepackage{longtable}
\usepackage{multirow}
\usepackage{ulem}

\usepackage{listings}

\usepackage{hyperref}
\usepackage{url}

\usepackage{xspace}

\newcommand{\ME}{M$_{\oplus}$\xspace}

\newcommand{\RE}{R$_{\oplus}$\xspace}

\newcommand{\vinf}{$v_{\infty}$}

\newcommand{\vesc}{$v_{esc}$}

\title{Bifurcation in the history of Uranus and Neptune: the role of giant impacts}
\author[C. Reinhardt, A. Chau, J. Stadel and R. Helled]{Christian Reinhardt$^1$\thanks{christian.reinhardt@ics.uzh.ch},
                                    Alice Chau$^1$,
								    Joachim Stadel$^1$,
									Ravit Helled$^1$\\
									\\
                                    $^{1}$Institute for Computational Science, University of Zurich, Winterthurerstrasse 190, 8057 Zurich, Switzerland	
									}

\date{Accepted XXX. Received YYY; in original form ZZZ}

\pubyear{2019}

\begin{document}
\label{firstpage}
\pagerange{\pageref{firstpage}--\pageref{lastpage}}
\maketitle

\begin{abstract}
Despite many similarities, there are significant observed differences between Uranus and Neptune: while Uranus is tilted and has a  regular set of satellites, suggesting their accretion from a disk, Neptune's moons are irregular and are captured objects.
In addition, Neptune seems to have an internal heat source, while Uranus is in equilibrium with solar insulation.
Finally, structure models based on gravity data suggest that Uranus is more centrally condensed than Neptune. 
We perform a large suite of high resolution SPH simulations to investigate whether these differences can be explained by giant impacts. 
For Uranus, we find that an oblique impact can tilt its spin axis and eject enough material to create a disk where the regular satellites are formed. Some of the disks are massive and extended enough, and consist of enough rocky material to explain the formation of Uranus' regular  satellites. 
For Neptune, we investigate whether a head-on collision could mix the interior, and lead to an adiabatic temperature profile, which may explain its larger flux and higher moment of inertia value. 
We find that massive and dense projectiles can penetrate towards the centre and deposit mass and energy in the deep interior, leading to a less centrally concentrated interior for Neptune. 
We conclude that the dichotomy between the ice giants can be explained by violent impacts after their formation.

%This is a simple template for authors to write new MNRAS papers.
%The abstract should briefly describe the aims, methods, and main results of the paper.
%It should be a single paragraph not more than 250 words (200 words for Letters).
%No references should appear in the abstract.
\end{abstract}


%%%%%%%%%%%%%%%%%%%%%%%%%%%%%%%%%%%%%%%%%%%%%%%%
% Introduction
%%%%%%%%%%%%%%%%%%%%%%%%%%%%%%%%%%%%%%%%%%%%%%%%
\section{Introduction}
Uranus and Neptune are the outermost planets of our solar system, located at a distance of 19.1 and 30.1 au from the Sun, respectively. 
Their similar masses (14.5~\ME and 17.1~\ME), mean densities (1.27~g cm$^{-3}$ and 1.64~g cm$^{-3}$), and large radial distances  from the Sun suggest that they form their own class of planets within the solar system, distinct from the inner terrestrial planets and the gas giants. 
At present, there are various efforts to design dedicated space missions to these planets which makes them prime objects for scientific investigations. 
\par

While Uranus and Neptune are often referred to as ice giants because of their mean densities, 
their actual water abundances are unknown (e.g. \citealt{Podolak2012, Helled2011}). 
In fact, there are still large uncertainties regarding their bulk compositions and internal structures. 
The fact that their temperature profiles could differ from adiabatic ones, that their interiors can consist of composition gradients and/or boundary layers, and that their rotation periods and shapes are not well determined, add additional complexity to structure models \citep{Helled2011, Nettelmann2013}.

Although they have
similar masses and sizes, there are crucial differences between the two planets. One prominent example is the large obliquity of Uranus: the rotational axis of the planet as well as its five regular moons is tilted by $\sim97$ degrees (retrograde) with respect to the solar plane, which is unique in our solar system. 
Uranus' five satellites are on regular orbits suggesting that they formed in a circumplanetary disk. 
On the other hand, Neptune’s largest moon, Triton, is in a very inclined orbit, and therefore is likely to be captured (e.g. \citealt{McKinnon1995}, \citealt{Agnor2006}). Neptune's outer small moons also seem like captured Trans-Neptunian and/or Kuiper belt objects.  
In addition, Uranus seems to be in thermal equilibrium with solar insulation while Neptune's thermal flux is about one order of magnitude larger \citep{Pearl1991}. An adiabatic interior is hence a reasonable assumption for thermal evolution models for Neptune, while for Uranus it suggests that either Uranus has cooled much faster than Neptune or that its heat is still stored within its interior and something prevents it from being effectively transported. If the heat is still trapped in Uranus' deep interior, it could be a result of the existence of a boundary layer and/or composition gradients that inhibit efficient convection within the planet (e.g., \citealt{Nettelmann2016, Podolak2019, Vazan2019}). Indeed thermal evolution models as well as alternative structure models show that an adiabatic cooling/temperature profile is appropriate for Neptune but not for Uranus \citep{Fortney2011,Nettelmann2016,Podolak2019}. Finally, structure models based on the available gravity data ($J_2, J_4$) suggest that Uranus is more centrally condensed than Neptune. This is somewhat consistent with the idea that Neptune is more homogeneously mixed (due to convection) while Uranus consists of more distinct layers, and possibly, a larger core \citep{Podolak2012}. 

It is possible that the ice giants shared a common formation path while giant impacts (GIs) occurring shortly after their formation have given them their distinct properties \citep{Stevenson1986, Podolak2012}. 
An oblique impact with a massive impactor could not only significantly alter Uranus' spin \citep{Safronov1966}, but could also eject enough material 
to form a disk where its regular moons are formed.  
An oblique impact typically does not affect the planetary internal structure, 
so any composition barrier that inhibits convection is expected to remain. On the other hand, Neptune could have experienced a head-on collision which led to a more mixed interior. 

While \citet{Podolak2012} investigated whether giant impacts could lead to some of the observed differences between Uranus and Neptune, the calculations were limited to the motion of the impactors through the planetary envelope and could only track the energy and angular momentum deposition. 
Previous studies using full 3D hydro-simulations focused solely on Uranus. \citet{Slattery1992} (S92) performed Smoothed Particle Hydrodynamics (SPH) simulations and showed that an impactor with a mass $> 1$\ME with an  impact velocity sightly above the mutual escape velocity could produce Uranus' rotation rate. 
Some of the simulations also produced a circumplanetary disk due to the tidal disruption of the impactor. The resulting disk was massive enough (about 1\% - 3\% of the total colliding mass) but too compact
(only a few Uranian radii) 
to readily explain the formation of the outer satellites \citep{Canup2000}. The low resolution of a few thousand particles did not allow a detailed analysis of the planetary internal structure, composition, and orbiting material. 

\citet{Kegerreis2018} (K2018) revisited this scenario with SPH simulations using a similar code with different equations of state (EOS) to model the materials and significantly higher resolutions ($10^5$ to $10^6$ particles). While they found a general agreement with S92, with the significantly higher resolution, the interior of Uranus and the orbiting material were resolved.
The collisions lead to deposition of shocked material from the impactor into the planet's interior, forming a hot, high-entropy layer. 
It was also found that projectiles up to 3~\ME are tidally disrupted and efficiently deposit rocky material in orbit, which differs from the findings of S92, probably due to the improved resolution.

K2018 also performed the first 3D simulations on atmospheric loss in giant impacts finding that $> 90\%$ of the atmosphere remains bound to the planet, but depending on the impact conditions, can be outside of the Roche limit which affects the conditions for satellite formation. 
In a following paper, \citet{Kegerreis2019} revisited the scenario with higher resolution simulations. The results were in general agreement with their earlier work, and revealed more 
information regarding the composition of the orbiting material, and the tidal disruption of the impactor's core in grazing collisions. 

Neptune, on the other hand, has received less attention. \citet{Podolak2012} performed 1D calculations of impacts on Neptune's envelope but their computations did not include a detailed modelling of hydrodynamic effects. To our knowledge, there are no 3D hydro-simulations that investigate how an impactor of several \ME would affect Neptune's interior. Such massive bodies can in principle deposit mass and energy deep in the planet's interior, and therefore are ideal candidates to study the effects of impacts on Neptune's long-term thermal evolution.

In this paper we present an extensive set of state-of-the-art GI simulations for both Uranus and Neptune using a common simulation framework, and featuring high resolution SPH calculations with low--noise initial conditions, in order to investigate whether the dichotomy between the planets can be explained by GI. 
Our paper is structured as followed: in Section~\ref{section:methods} we present the numerical method and the equations of state used in our simulations. We also discuss the pre-impact planets and how the initial conditions are built. In Section~\ref{section:uranus} and Section~\ref{section:neptune} we present the results for Uranus and Neptune, respectively. 
A summary and the discussion of the result as well as an outlook for future research are presented in Section~\ref{section:discussion}.

%%%%%%%%%%%%%%%%%%%%%%%%%%%%%%%%%%%%%%%%%%%%%%%%
% Methods, Initial conditions
%%%%%%%%%%%%%%%%%%%%%%%%%%%%%%%%%%%%%%%%%%%%%%%%
\section{Methods}
\label{section:methods}
The impact simulations are performed using the SPH code \textsc{Gasoline} \citep{Wadsley2004} with the modifications for planetary collisions described in \citet{Reinhardt2017}. A free surface treatment, in combination with \textsc{ballic}, allows stable models to be generated without wasting time on relaxation prior to the impact calculation. We use both standard SPH \citep{Monaghan1992} and our fully entropy conserving ISPH algorithm. The use of the Wendland $C_2$ kernel \citep{Dehnen2012} avoids the numerical clumping instability that can occur when using the standard cubic spline kernel.

\subsection{Density correction at material interfaces}
\label{subsection:densitycorrection}
Standard SPH fails in capturing discontinuities \citep{Agertz2007}, e.g., encountered at the core-mantle boundary of a planet, resulting in severe 
over or under 
estimate of the particle's density at the interface. This is problematic since it affects the model's stability, requires careful relaxation, and  also causes a gap at the interface (e.g. \citealt{Canup2001a}) which inhibits mixing and at the same time smooths out discontinuities. 
For rather cold models (low thermal energy) particles of the lower density material can enter unphysical states affecting the stability of the simulation. This is even more critical when ISPH is used, since in this method the particles are required to be above the minimum energy state of the material. 

Most prior work on capturing discontinuities with SPH (e.g., \citealt{Price2008}, \citealt{Read2012},  \citealt{Hosono2016}) required drastic changes to the algorithm. Here we present a different, simpler method that overcomes most of the difficulties encountered when applying such algorithms to a non-ideal EOS such as the Tillotson EOS 
used in this work. 
In order to build particle representations of giant planets, \citet{Woolfson2007} suggested to correct the density at a material interface by assigning particles of different material a different weight in the SPH density sum: 
\begin{equation}
    \rho_i = \sum_{j \in NN} {f_{ij} m_j W_{ij}}
\end{equation}
where
\begin{equation} 
    \label{eqn:fij}
    f_{ij} = \frac{\rho_i\left(P, T\right)}{\rho_j \left(P, T\right)}, 
\end{equation}
assuming that the pressure and temperature on the kernel is approximately constant. In Woolfson's paper this modification was applied to equilibrium models of giant planets with a four-layer structure including an iron core, a rocky mantle, an "ice" layer and a H-He gaseous envelope. Since the models were static, i.e., were not dynamically evolved in an SPH code, the pressure and temperature of each particle was known from the equilibrium calculations which substantially simplified the density correction.  

In impact simulations the pressure and temperature are {\it a priori} unknown and are calculated based on the  particle's density, which is severely over-- or under--estimated at the interfaces, and therefore the above approach needs to be modified in order to be applicable for impact simulations. 
One way to obtain good pressure and temperature estimates is to calculate the kernel averaged mean, 
which is expected to be nearly constant, thus cancelling out the large fluctuations at the interface. 
We obtain the best results when doing a simple arithmetic mean. 
Using a geometric mean results in more accurate estimates since very large values contribute less but can cause overflow errors when large pressures and temperatures are involved, e.g., due to shock compression during the impact. The resulting mean pressure and temperature are then used to determine the coefficients in equation (\ref{eqn:fij}) and correcting the density (see Appendix~\ref{appendix:interf_correction} for details). Since the fundamental SPH equations remain unchanged in this approach, the conservation properties of the method are not affected and the method can be implemented in any existing code without major changes. Note that the EOS only enters via the pressure and temperature estimate so the method does not explicitly depend on the choice of EOS. Therefore this method provides a very flexible tool for modelling contact discontinuities in impact simulations.

Our algorithm for the SPH density estimator at material interfaces is summarised as follows:
\begin{enumerate}
    \item Smooth the particle's (uncorrected) densities $\rho_i$ using the normal SPH density estimator
    \item Use $\rho_i$ and the internal energy $u_i$\footnote{Note that a particle's internal energy in SPH is not a smoothed quantity and therefore does not require any correction.} to obtain $P_i$ and $T_i$ for each particle from the EOS
    \item Calculate Kernel average $\overline{P}_i$ and $\overline{T}_i$ for all particles with a neighbour of differing material (an interface particle)
    \item Determine the correction factors $f_{ij} \left( \overline{P},  \overline{T} \right)$
    \item Re-smooth the density of interface particles according to equation (\ref{eqn:fij})
    \item Proceed with the usual SPH algorithm
\end{enumerate}

\begin{figure*}
	\centering
	\includegraphics[keepaspectratio,width=\textwidth]{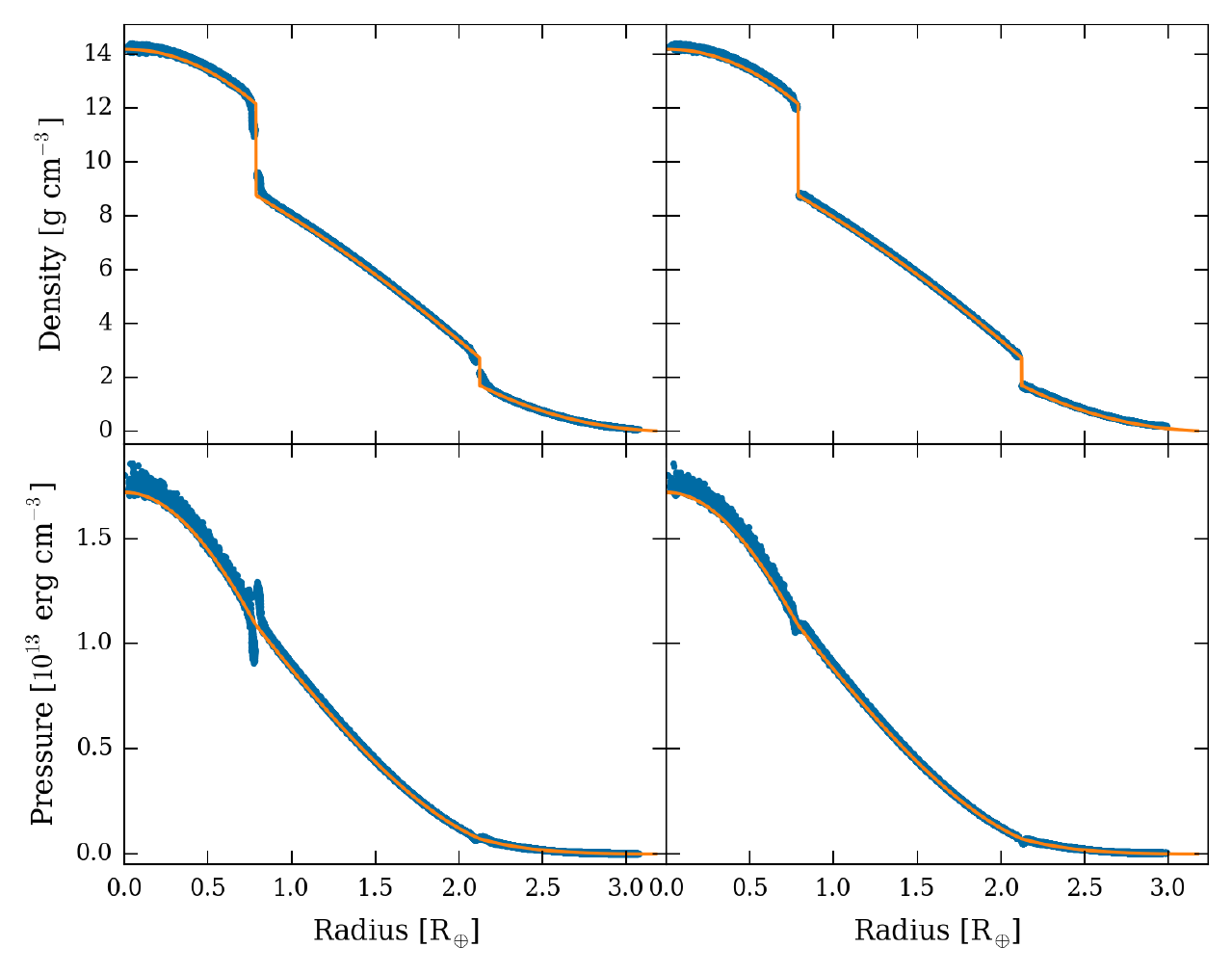}

	\caption{The radial density (top) and pressure (bottom) profiles of a 11.5~\ME pre-impact target (orange line) sampled with $10^6$ particles relaxed for 26~h in simulation time with classic SPH (left) and using the material interface treatment presented in this work (right).
	The blue dots show the particles density and pressure (which is used to calculate the pressure forces in the simulations).
	The left two plots demonstrate that standard SPH fails at capturing the material interfaces, and leads to a pressure blip. With our interface treatment all discontinuities are modelled correctly, and the resulting pressure is continuous across the interfaces.}
	\label{fig:model_target}
\end{figure*}

When we apply the above algorithm to static models of proto-Uranus (or proto-Neptune) we find that the SPH density estimator perfectly follows the imprinted profile (Figure~\ref{fig:model_target}) and the pressure blip at the interface completely vanishes.

%%%%%%%%%%%%%%%%%%%%%%%%%%%%%%%%%%%%%%%%%%%%%%%%
% Fig: Radial density and pressure profile 
%%%%%%%%%%%%%%%%%%%%%%%%%%%%%%%%%%%%%%%%%%%%%%%%

\subsection{Equilibrium models}
\label{subsection:models}
The SPH representation of the target and the impactor were obtained as described in \citet{Reinhardt2017}. In order to build differentiated bodies with multiple materials the procedure was slightly modified. Rather than solving the structure equations iterating for different values of the density and internal energy, which usually requires a good initial guess for convergence, we build a grid of models varying the density and internal energy at the core. 
The model that has the desired density and energy at the surface, and best matches the required mass (within $10^{-6}$) is used to build the particle representation of the colliding bodies. 
To properly capture the material boundaries, the particles are distributed on each material layer (core, mantle and envelope) separately. Then we iterate over all of them until the distribution converges (see \citealt{Reinhardt2017} for details). 
The particle mass is taken to be the layer's total mass divided by the number of particles in that layer. In principle this should result in equal mass particles, as required to maintain stability in SPH (e.g., \citealt{Mastropietro2005}). Due to constraints from the \textsc{healpix} grid \citep{Gorski2005} the particle number can vary, however, resulting in slightly varying particle mass ratios. For all models this mass ratio is always very nearly 1:1 and thus does not affect the numerical stability of the simulations.

Since the pre-impact composition of Uranus and Neptune are poorly constrained we follow \citet{Nettelmann2013} and model the planetary interior with three distinct layers: a rocky core composed of silicates, an inner water envelope (hereafter,  ice mantle), and an outer gaseous H-He envelope.
The total colliding mass (target and impactor) is set to Uranus' and Neptune's observed values of 14.5 and 17.1~\ME, respectively.
For our simulations we use the Tillotson EOS \citep{Tillotson1962} to model the heavy elements, granite \citep{Benz1986}  for the rock  and water ice \citep{Benz1999}. The Tillotson EOS is a relatively simple, analytic EOS and was developed to model hyper-velocity impacts. It has been used in many prior studies on GI due to its excellent ability to model shocks and to cover the wide ranges of densities and temperatures expected in such violent collisions.
Although the Tillotson EOS lacks the representation of phase transitions and mixed phases,  
it agrees well with experiments  (e.g., \citealt{Brundage2013}) and faithfully reproduces shocks which is crucial for modelling hyper-velocity impacts.
The planet's H-He envelope is modelled using an ideal gas EOS
with the mean molecular weight set to $1.0$ times the mass of a hydrogen atom in order to have a more physical behaviour of the gaseous layer (i.e., the temperature at the discontinuity between the inner and outer envelope is closer to more realistic models). While such a simple EOS is inappropriate for large densities (and corresponding large pressures), it provides a simple  description of a low-density gas. We plan to incorporate a more physical EOS for H-He in future research. 
Since an ideal gas is compressible without limit, in some cases, the inferred density at the mantle-envelope boundary can have high values that lead to unphysical models.
This problem does not occur if very cold models, e.g., with surface temperatures below 50 Kelvin, are avoided. 
Since Uranus and Neptune have surface temperatures are above this value, and expected to be hotter shortly after their formation, none of our models are affected by this issue.

The pre-impact targets are assumed to consist of a 10\% (by mass) rocky core surrounded by an ice mantle and 2~\ME H-He envelope.
The resulting bodies are in relatively good agreement with predictions from interior models of Uranus and Neptune that use more sophisticated EOS.
They contain more than 70\% heavy elements, have a discontinuity (mantle-atmosphere boundary) at about 70\% of the planet's radius, their normalised moment of inertia (MOI) are between 0.21 and 0.22 and the ice-to-rock ratio is above the solar value of 2.7 (\citealt{Helled2011}, \citealt{Nettelmann2013}).
However, interior models as well as observations of the ice giants suggest that their  H-He  atmospheres are significantly enriched in heavy elements, and this characteristic is not included in our models since we use an ideal gas EOS for H-He. 
Given that the internal structure of proto-Uranus and proto-Neptune are unknown, the shortcomings of our numerical method can be considered acceptable. We focus on the investigation of the trends and the type of impacts that can affect the planetary internal structure.
Clearly, our findings presented are affected by the assumed pre-impact planet's composition, which is unknown and in principle could be rather different from our models. However, given the large uncertainties on the inferred composition of Uranus and Neptune from interior models, our assumed internal structure models are acceptable.  
Nevertheless, we also consider impacts on an extreme case of a solid initial proto-Uranus composed of 10\% rock and 90\% ice in Appendix \ref{appendix:two_component_models} in order to check the sensitivity of our findings to the assumed EOS. 
A detailed investigation of the effect of the assumed target's internal structure and composition on the GI simulation results is clearly desirable but is beyond the scope of this paper, and we hope to address it in future research. 
For the projectiles, we consider three different compositions including pure-rock, pure-ice, and a differentiated impactor composed of 12\% rock and 88 \% ice (similar to the target's composition, hereafter, "differentiated") in order to check the sensitivity of the results to the impactor's composition.
The ice-to-rock ratio of the differentiated impactors is a free parameter and can have a large range.
Clearly future simulations should consider other compositions, especially as several objects in the outer part of the solar-system, like Pluto, are found to be rock-dominated \citep{McKinnon2017}.
The pure rock or ice impactor of several Earth masses, as considered in this study, are extreme cases and should be taken as end-members for the possible composition of the impactor.
We also consider three values for the impactor's mass of 1, 2 and 3~\ME. 
The target's mass is then adjusted accordingly, so that the total colliding mass matches the masses of Uranus and Neptune for the merging collisions.
For a given resolution of the target, the number of particles sampling the impactor is adapted, so that all particles have (almost) the same mass. For example, a 12.5~\ME proto-Uranus represented with $10^5$ particles collides with a 2~\ME impactor sampled with 1.6$\times10^4$ particles.

\subsection{The Simulation Suite}
\label{subsection:ic}
We assign no initial rotation to the target or the impactor prior to the collision. Since the pre-impact spin is unknown and GI substantially alter the planet's angular momentum, this assumption is reasonable in the context of our study. However, if one aims to determine the origin of the projectile or further constrain the impact conditions, the pre-impact state of the target has to be considered. 
In all merging simulations we set the relative velocity at infinity \vinf=5~km s$^{-1}$ leading to impacts that are slightly above the mutual escape velocity of the system, i.e., the normalised impact velocity is $v_{imp}$/$v_{esc}$ $\sim$ 1.03 for all impactor masses and compositions.
The displacement of the target and the projectile at the impact is determined from the  impact parameter $b$, where $b=0$ is a head-on collision and $b=1$ means that the bodies do not interact.
 This property is somewhat more intuitive than the total angular momentum to describe the initial conditions.
In the case of Uranus, we vary the impact parameter between 0.1 and 0.9 for all impactor masses and compositions.
For the Neptune case we limit the impact parameter to $\leq0.5$, as more grazing collisions are unlikely to lead to penetration to the deep interior.
Prior to the impact both bodies are slightly more separated than the sum of their radii and assigned an impact velocity $v_{imp}^2 = v_{esc}^2 + v_{\infty}^2$, where \vesc\xspace is their mutual escape velocity. 
Collisions at velocities close to the mutual escape velocity are the most likely outcome of a gravitational interaction  between two bodies.
The resulting impact velocities of 18 to 20~km~s$^{-1}$ are larger than Uranus' or Neptune's orbital velocities (which are $\sim$ 6~km~s$^{-1}$) and are therefore at the upper end of the expected relative velocities. 

A third class of impacts we investigate are hit-and-run collisions (HRC) between Uranus and a twin planet of the same mass.
Since such collisions by definition lead to little accretion or erosion the target, for this case the target's mass is set to that of Uranus.
For the HRC the impact velocity ranges from 2~$v_{esc}$ to 4~$v_{esc}$ depending on the specific impact conditions \citep{Leinhardt2012}. Since it is found that such impacts deposit substantially less angular momentum in the planet, for this scenario we also considered initially rotating models, where the pre-impact planetary rotation varies from 20~h to 30~h
(see Section~\ref{subsection:hit-and-run-collision} for details).  

In order to cover a large parameter space of collisions, we use a moderate resolution for the first suite of simulations, and model the target with $10^5$ particles. 
Such simulations require less than one day per collision on a single node allowing us to investigate various impact angles, impact velocities, impactor compositions, and different numerical parameters (e.g., resolution, treatment of boundaries and discontinuities and viscosity limiter). We then successively increase the resolution to $10^6$ and in some cases to $5\times10^6$ particles in order obtain a more detailed picture of the post-impact target and the orbiting material, and to investigate how the different numerical parameters affect convergence.
All simulations are run for at least 80 hours in simulation time. However, in some cases the grazing collisions ($b$ > 0.8) required substantially more time because the impactor survives the initial impact and re-impacts within 8 days.
The full suite of simulations required 8'000'000 CPU--hours\footnote{222'000 node--hours on the Piz Daint "multi--core" partition at the Swiss National Supercomputing Center in Lugano, Switzerland.} and is summarised in Table 1. 

\subsection{Analysis}
\label{subsection:analysis}
All impact simulations result in one or two final post-impact bodies, the target, and in the case of very grazing or 
HRC, an impactor remnant. 
In order to distinguish them from the surrounding ejecta we use \texttt{SKID}\footnote{The source code is available at: \url{http://faculty.washington.edu/trq/hpcc/tools/skid.html}.} \citep{Stadel2001} to determine coherent, gravitationally bound clumps of material.
For our analysis we use the following parameters: the number of smoothing neighbours \textit{nSmooth} is set to 400, 800 and 1600 for the $10^5$, $10^6$ and $5 \times 10^6$ particle simulations, respectively, and the linking length \textit{tau} is 0.06~\RE. 
We find that the results are insensitive to large variations (one order of magnitude) of the parameter \textit{tau}. However, it is important to choose at least several hundred smoothing neighbours in order to reduce noise in the density estimate and prevent the algorithm from finding artificial subgroups.

This procedure leads to a central dense region we refer to as \textit{planet} surrounded by an envelope of gravitationally bound, low-density material. This orbiting material can be further divided into an extended \textit{atmosphere} and a \textit{circumplanetary disk}. In this work, we distinguish the disk from the rest of the orbiting material using the algorithm of \citet{Canup2000}. This algorithm first determines the particles that belong to the planet using Uranus' or Neptune's mean density. Then all the particles that are gravitationally bound to the planet are found. Depending on their angular momentum (with respect to the planet) the bound particles are either added to the planet or considered as part of the disk. Using the updated estimate of the planet's mass the algorithm iterates until the masses converge (see \citealt{Canup2000} for further details). 

%%%%%%%%%%%%%%%%%%%%%%%%%%%%%%%%%%%%%%%%%%%%%%%%
% Fig: Rotation period (density intervals) 
%%%%%%%%%%%%%%%%%%%%%%%%%%%%%%%%%%%%%%%%%%%%%%%%
\begin{figure}
	\centering
	\includegraphics{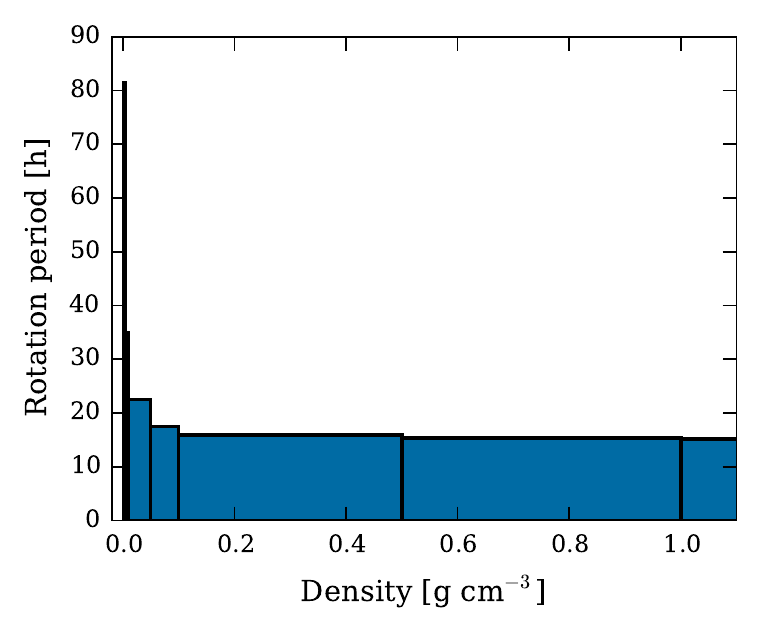}

	\caption{\textbf{Uranus' rotation period for different cut-off densities}. Uranus' rotation period after colliding with a 2~\ME differentiated impactor ($b=0.2$, \vinf=5 km s$^{-1}$) when different cut-off densities are considered (using $5\times10^6$ particles). The rotation period of each layer is inferred from its angular momentum and moment of inertia values 26 hours after the impact as described in Section \ref{subsection:analysis}. Except for the most outer low-density layer, the inferred rotation periods are similar, are nearly constant, and are found to be in good agreement with other methods (see text for further details).}
	\label{fig:uranus_rot_period_vs_density}
\end{figure}

The post-impact rotation period is determined as follows. First, we define the planet as described above, then 
we divide the SPH particles into spherical bins, and calculate the average angular momentum of each bin in order to reduce noise inherent to SPH. We can therefore infer a continuous radial angular momentum profile to which we fit a solid-body rotation from:
\begin{equation}
    L = m \omega r^2, 
\end{equation}
where the rotation period is $P=\omega/2\pi$. 
In order to test the sensitivity of the result on the method, we independently determine the rotation period from: 
\begin{equation}
    L = I \omega,
\end{equation}
where $I$ is the body's moment of inertia, and $L$ is the total angular momentum of all particles. We also calculate the rotation period from the median of the particle's angular velocities following \citet{Kegerreis2018}. 
Overall, the various methods are in good agreement. Our method diverges if one also accounts for the low density orbiting material that deviates from solid body rotation.
However, an analysis of this material (Figure~\ref{fig:uranus_rot_period_vs_density}) shows that the rotation periods in different density layers remain similar. Only the outer most layer rotates substantially slower.

While the rotation period can change with time, e.g., due to cooling and contraction of the planet, the angular momentum is conserved. As a result, angular momentum may seem to be  a more suitable quantity to describe the post-impact state and for comparison with Uranus. Since the orbiting low-density material contains a substantial fraction of the angular momentum, the result strongly depends on the definition of the "planet" (see the beginning of this Section) and the amount of disk material that is later reaccreted.

%%%%%%%%%%%%%%%%%%%%%%%%%%%%%%%%%%%%%%%%%%%%%%%%
% Uranus
%%%%%%%%%%%%%%%%%%%%%%%%%%%%%%%%%%%%%%%%%%%%%%%%
\section{Uranus}
\label{section:uranus}

The extreme tilt of Uranus' spin axis remains the most prominently compelling feature for a giant impact scenario.  Our simulations start with initially non-rotating bodies  such that the angle of the impact plane with respect to the Solar System's plane remains completely unspecified due to symmetry. This means that any of our collision simulations\footnote{Except in the Hit-and-Run (HRC) case where we also consider cases with initial pre-impact spin of the target.} can reproduce the desired value of the planet's obliquity.  While the pre-impact rotation, which is determined by the formation process is unknown, it is expected to be small \citep{DonesTremaine93}. 
Following \citet{Slattery1992}, we focus on impact conditions and impactor compositions that can reproduce Uranus' rotation period of 17.24~h from a  non-rotating pre-impact Uranus,  
as well as the formation of a circumplanetary disk.
We also investigate the internal structure and atmospheric composition of Uranus after the impact.
\par

\begin{figure}
	\centering
	\includegraphics{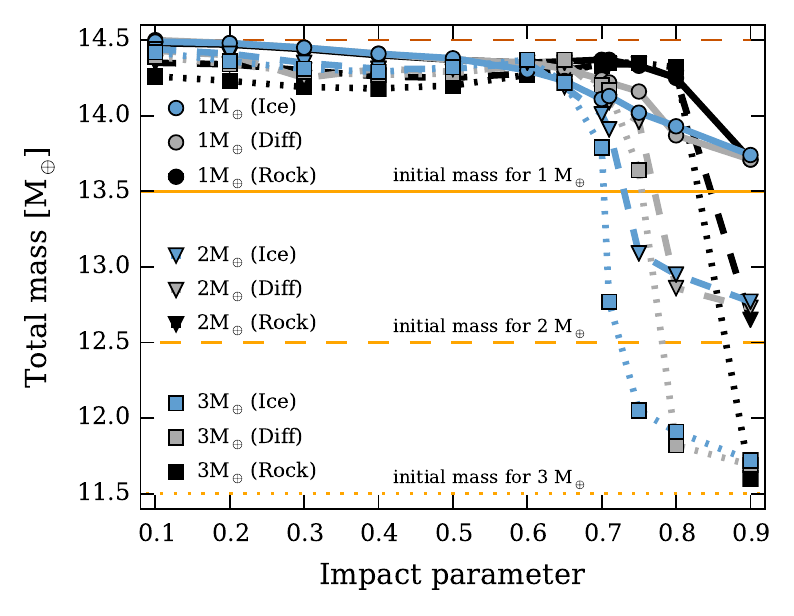}

	\caption{\textbf{Post-impact total bound mass of the Uranian system (planet+envelope+disk) for different impactor masses and compositions.}
	The total mass colliding is set to Uranus' observed value (14.5~\ME, dashed red line). The different symbols represent different impactor masses	(circle: 1~\ME, triangle: 2~\ME, square: 3~\ME) and the colours correspond to the impactor's composition (blue: ice, grey: differentiated, black: rock).
	The orange lines show the planet's initial mass which depends on the impactor's mass (solid: 1~\ME, dashed: 2~\ME and dotted: 3~\ME).
	In all cases the target is represented with $10^5$ particles (see Section \ref{subsection:ic} for details).
	Most of the impactor is accreted for $b<0.71$. For larger impact parameters the impactor can survive the collision and escape the system (HRC) resulting in little transfer of mass and angular momentum. Since icy and differentiated impactor enter the hit-and-run regime before rocky ones, they are less efficient at depositing mass in very grazing collisions (see Section~\ref{section:uranus} for details).}
	\label{fig:uranus1_mass}
\end{figure}

Figure~\ref{fig:uranus1_mass} shows the total bound mass around Uranus (including the disk) as a function of the impact parameter for various impactor masses and compositions. It is found that collisions with impact parameter up to $b\sim0.7$ lead to an-almost complete merging of the impactor and the target. This is valid for all impactor masses and compositions we consider. 
More massive impactors are more erosive as the initial targets are less massive and thus have a lower gravitational binding energy. Such impactors also lead to a larger envelope because the collision is more energetic and more material
 is (partially) vaporised. 
For larger angles, the impactor can survive the collision and leave the system, with almost no mass transferred to the target for collisions at $b\sim0.8-0.9$.
\par

We note that the lower-density impactors enter the 
HRC 
regime
for lower impact parameters than the denser ones for a given
 impactor mass and 
and impact velocity.
For a given impactor's mass, the lower-density impactors have larger sizes, and hence more of the material "misses" the target. 
In other words, the denser the impactor, the larger 
the mass fraction that interacts with the target during the collision. Since the mass (and momentum) stripped from the impactor during the encounter with the planet is approximately the initially overlapping mass, rocky impactors lose more of their initial momentum than icy ones, and tend to be more gravitationally bound after the impact.
This interpretation is supported by test simulations in which the same mass fraction of the impactor
interacts with the target, where we see that the outcome of the collision does not depend on the impactor's mean density.

%%%%%%%%%%%%%%%%%%%%%%%%%%%%%%%%%%%%%%%%%%%%%%%%
% Fig: Total bound mass after the impact 
%%%%%%%%%%%%%%%%%%%%%%%%%%%%%%%%%%%%%%%%%%%%%%%%

The exact mass that is accreted by the target and the location where it is deposited within the planet depend on the impactor-to-target mass ratio, but also the impactor's composition. Typically rocky impactors deposit more mass in the inner part of the planet since they are denser and penetrate deeper. As a result,  most of the rocky material is deposited above the target's core.
Only very grazing collisions of differentiated/rocky impactors 
can deposit rocky material in the planetary outer envelope or the disk because the projectile survives the first impact and is later tidally disrupted. In extreme cases the impactor can reach a distance of up to $\sim200$~\RE before colliding a second time with the planet.
The tidal disruption of the impactor leads to large streams of material that are later accreted by the planet.
In the case of a differentiated impactor, its core is also eroded and forms small clumps that are accreted by Uranus. 
These streams of in-falling material are observed for all resolutions. 
However, the disruption of the impactor's core can only be resolved with $> 10^6$ particles with classic SPH.
When the interface correction proposed in this paper is applied, core erosion is already observed in the lower resolution simulations, probably due to the reduced artificial surface tension at the core-mantle boundary (see Appendix~\ref{appendix:box_test} for details). 

Pure-ice impactors, on the other hand, remain in the target's upper envelope and atmosphere and cannot penetrate to the planet's deep interior.
This outcome is independent of the assumed impactor's mass or the impact angle. Differentiated impactors result in an intermediate outcome. The rock ends up in the planet's interior and ice in the outer layers.
Almost head-on collisions ($b < 0.4$) can also deposit ice from the impactor closer to the planet's core, but this never happens in the case of a pure-ice impactor. 

\subsection{Rotation period}
\label{subsection:rotation_period}
\begin{figure}
	\centering
	\includegraphics{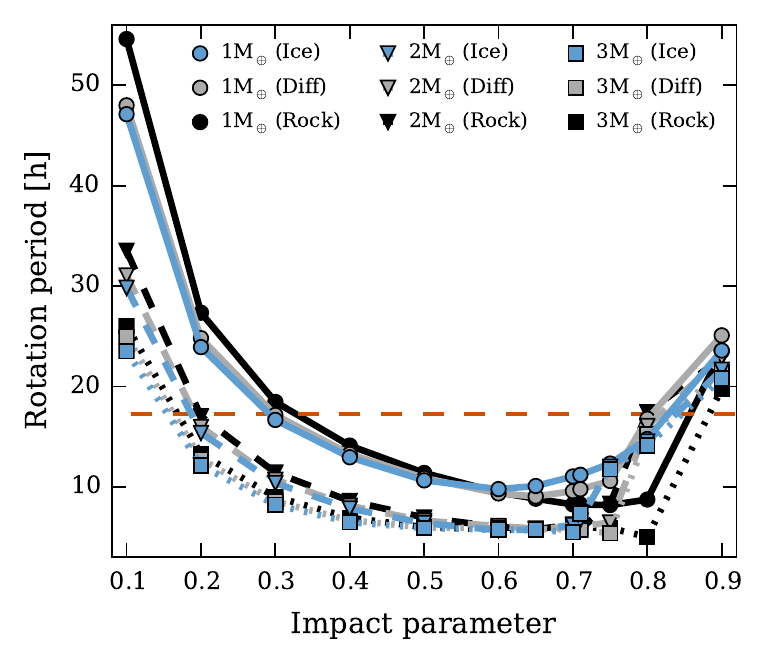}

	\caption{\textbf{Uranus' post-impact rotation period for different impactor masses and compositions.} Uranus' current rotation period of (17.24~h) is shown with a dashed 
	red 
	line. The different symbols and colours correspond to different impactors masses and compositions as indicated in the legend. 
    The initial conditions are set as described in Section \ref{subsection:ic}, with a non-rotating proto-Uranus (resolved with $10^5$ particles) prior to the collision .
	Most of the collisions lead to a rotation period that is shorter than 17.24~h. Only almost head-on or very grazing collisions most be excluded as candidates to explain Uranus' spin.}
	\label{fig:uranus1_spin}
\end{figure}

In Figure~\ref{fig:uranus1_spin} we show Uranus' post-impact rotation period as a function of the impact parameter for different impactor masses and compositions.
Head-on collisions ($b<0.2$) cannot substantially alter the planetary spin. The rotation period decreases with increasing impact parameter, until a plateau is reached around $b \sim 0.5-0.7$. A turnover is observed for larger impact parameters when the impacts enter the HRC regime. 
Massive impactors have higher angular momenta and therefore lead to faster rotation. 
For the initial condition we consider, an increase of 1~\ME to the impactor's mass shortens the target's rotation period by a factor of $1/3$.

The impactor's composition also mildly affects the resulting rotation rate: pure-ice impactors transfer angular momentum to the target more efficiently than differentiated or rocky bodies because the icy bodies can only penetrate the target's outer layers while the denser objects reach deeper regions.
For $b>0.7$, the impactor mostly interacts with Uranus' atmosphere. While it is deflected from its original trajectory and loses some kinetic energy, a remnant 
of the projectile survives the collision.
The projectile can remain bound and is tidally disrupted or re-impacts during a following encounter. 
While the general trend agrees well with previous work (\citealt{Slattery1992} and \citealt{Kegerreis2018}) we find that also a 1~\ME impactor can reproduce Uranus' rotation. 
It should be noted that spin-orbit resonances \citep{Rogoszinski2019} as well as a hot high-entropy initial target \citep{Kurosaki2019} can reduce the required impactor mass.

%%%%%%%%%%%%%%%%%%%%%%%%%%%%%%%%%%%%%%%%%%%%%%%%
% Fig: Rotation period after the impact 
%%%%%%%%%%%%%%%%%%%%%%%%%%%%%%%%%%%%%%%%%%%%%%%%

%%%%%%%%%%%%%%%%%%%%%%%%%%%%%%%%%%%%%%%%%%%%%%%%
% Fig: Time evolution of the rotation period 
%%%%%%%%%%%%%%%%%%%%%%%%%%%%%%%%%%%%%%%%%%%%%%%%
\begin{figure}
	\centering
	\includegraphics{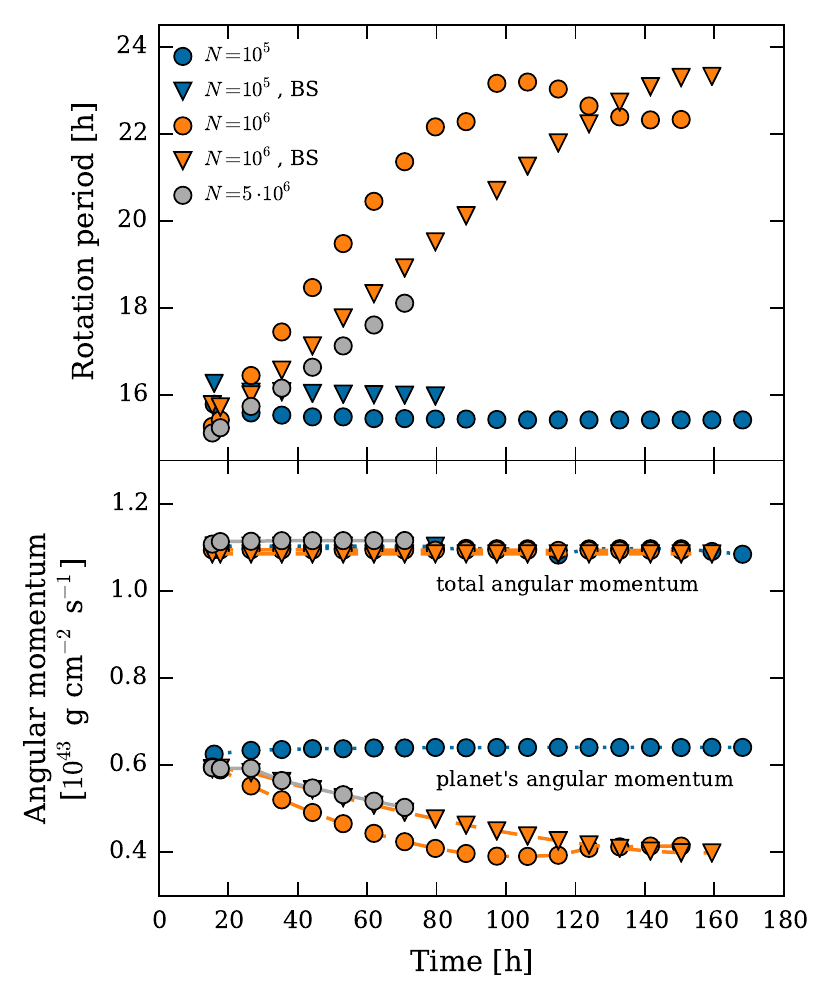}

	\caption{\textbf{The inferred rotation period and angular momentum for different resolutions versus time.}
	\textbf{Top panel:} The time evolution of Uranus' rotation period after the collision with a differentiated impactor of 2~\ME at $b$=0.2 with $v_{imp}=19.48$~km s$^{-1}$ for different resolutions $N$. For the low resolution simulation ($N=10^5$ particles, blue symbols) the rotation period converges quickly after the impact and remains constant over time, unlike for the higher resolution simulations (orange and grey symbols) where the planet's rotation period increases over time for simulations with (triangles) and without (circles) viscosity limiter. 
	\textbf{Bottom panel:} The total angular momentum (planet+envelope, continuous lines) and planet's angular momentum (dashed lines) for the same collision. 
	The total angular momentum is conserved in all cases but the planet's angular momentum is transferred to the envelope as time progresses for the high resolution simulations ($N=10^6$ and $5 \times 10^6$ particles) due excess artificial viscosity.
	Using a viscosity limiter (Balsara switch, triangles) decreases the angular momentum transfer but cannot remove it entirely (see Section \ref{subsection:rotation_period} for details).
	}
	\label{fig:uranus_rot_period_vs_time}
\end{figure}

We find that the inferred rotation period also depends on the simulation's resolution. 
Figure~\ref{fig:uranus_rot_period_vs_time} shows the time evolution of Uranus' rotation period after colliding with a 2~\ME differentiated impactor at $b=0.2$ and $v_{imp}=19.48$~km s$^{-1}$ using different resolutions. 
Simulations with $10^5$ particles lead to a constant rotation period that converges quickly after the collision.
For higher resolutions the rotation period initially agrees with the $10^5$ particle runs but is then increasing over time. 
While the total angular momentum is conserved in all cases (see Figure~\ref{fig:uranus_rot_period_vs_time}), there is a transport of angular momentum from the planet to the envelope
in the high resolution simulations, which increases the planet's rotation period over time.
Higher resolution simulations better resolve the differentially rotating flow in the upper mantle and atmosphere thus triggering unwanted artificial viscosity in this shearing flow. The Balsara switch \citep{Balsara1995} reduces artificial viscosity, and hence angular momentum transfer, in differentially rotating flows, but does not eliminate this effect 
entirely \citep{Cullen2010}. We find that further increasing the resolution, thereby reducing artificial viscosity, from $10^6$ to $5 \times 10^6$ particles lead to a slower decay of the rotation period; these (our highest resolution) simulations agree with the $10^6$ particle Balsara switch simulations.
Obtaining convergence in planetary rotations seems to require higher resolution and/or lower viscosity simulations and requires further investigation in the future.

\subsubsection{Envelope enrichment}
\label{subsection:envelope_enrichment}
\begin{figure}
	\centering
	\includegraphics{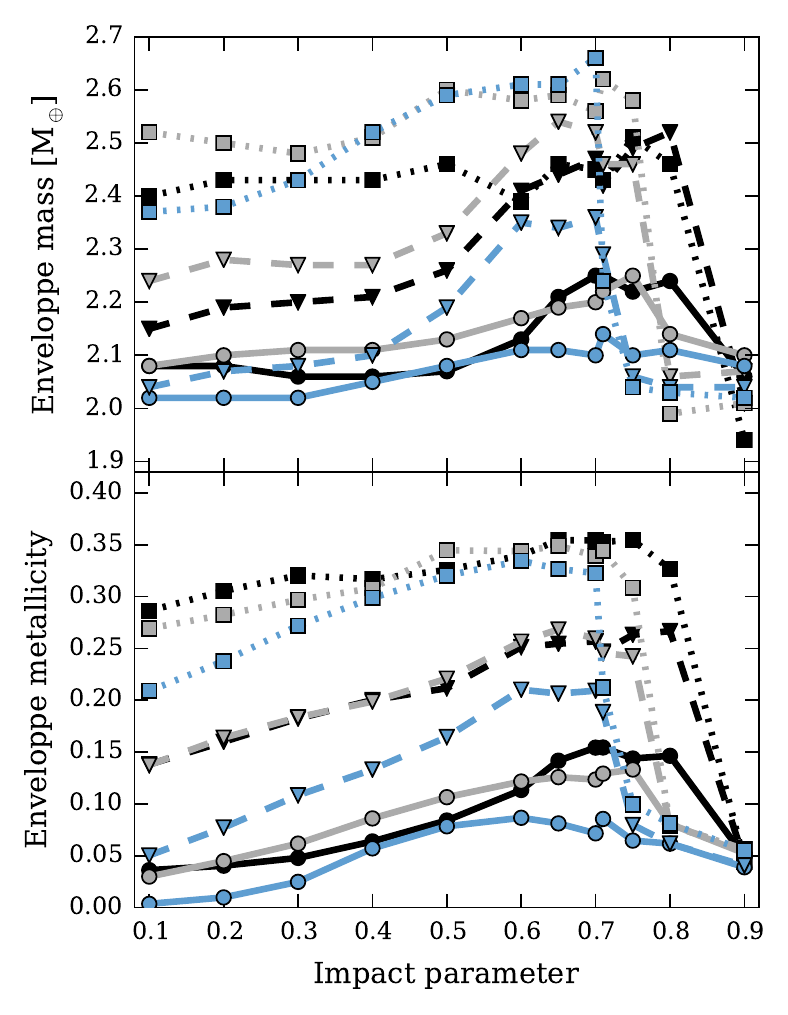}

	\caption{\textbf{The mass of Uranus' envelope and its metallicity after the collision.} The symbols represent different impactor masses (circle: 1~\ME, triangle: 2~\ME, square: 3~\ME) and the colours the  composition (blue: ice, grey: differentiated, black: rock). The target is resolved with $10^5$ particles.
	\textbf{Top panel:} The envelope's mass increases with increasing impactor mass because more energy is deposited in the planet and thus more material 
	(planet and impactor) is vaporised.
	Larger impact parameters lead to slightly more massive envelopes. Collisions with $b > 0.8$ are
	HRC 
	so little mass is added to the planet's envelope. For a 3~\ME granite impactor it is even partially eroded. In all collisions part of the primordial H-He envelope is ejected or temporarily captured by the escaping impactor. 
    \textbf{Bottom panel:} The inferred envelope's metallicity. More massive impactors result in higher  envelope metallicity. In all cases the envelope is enriched compared to its original pure H-He composition, except for the head-on collision with an ice projectile which does not affect the planet's mass and composition.
    }
	\label{fig:uranus1_envmetallicity}
\end{figure}

Figure~\ref{fig:uranus1_envmetallicity} shows the envelope's mass and inferred metallicity versus the impact parameter for different impactor masses and compositions.
We find that massive impactors vaporise more material in the collision. They produce heavier and more enriched envelopes. Grazing collisions ($0.5 < b < 0.8$) deposit more material in the envelope than head-on collisions ($b < 0.5$). In grazing collisions, the impactor is tidally stripped and the low density material remains in the envelope. Collisions with $b > 0.8$ are HRC, so little mass is added to the planet's envelope. For a 3~\ME rocky impactor the envelope is even partially eroded. We find that in all the collisions a fraction (up to 10\%) of the primordial H-He envelope is ejected, incorporated into the disk or escapes with the impactor. In addition, in all the cases, the planetary envelope is enriched with heavy elements (water/rock) compared to its original pure H-He composition (up to $\sim$35\% or 17.5 times the solar value when assuming $Z_{\odot}=0.02$).
This is consistent with structure models of Uranus and Neptune that infer high-metallicities in their atmospheres (e.g., \citealt{Helled2011}, \citealt{Nettelmann2013}).

%%%%%%%%%%%%%%%%%%%%%%%%%%%%%%%%%%%%%%%%%%%%%%%%
% Fig: Mass of the envelope 
%%%%%%%%%%%%%%%%%%%%%%%%%%%%%%%%%%%%%%%%%%%%%%%%

\subsubsection{Satellite disk formation}

%%%%%%%%%%%%%%%%%%%%%%%%%%%%%%%%%%%%%%%%%%%%%%%%
% Fig: Mass of the disk
%%%%%%%%%%%%%%%%%%%%%%%%%%%%%%%%%%%%%%%%%%%%%%%%
\begin{figure}
	\centering
	\includegraphics{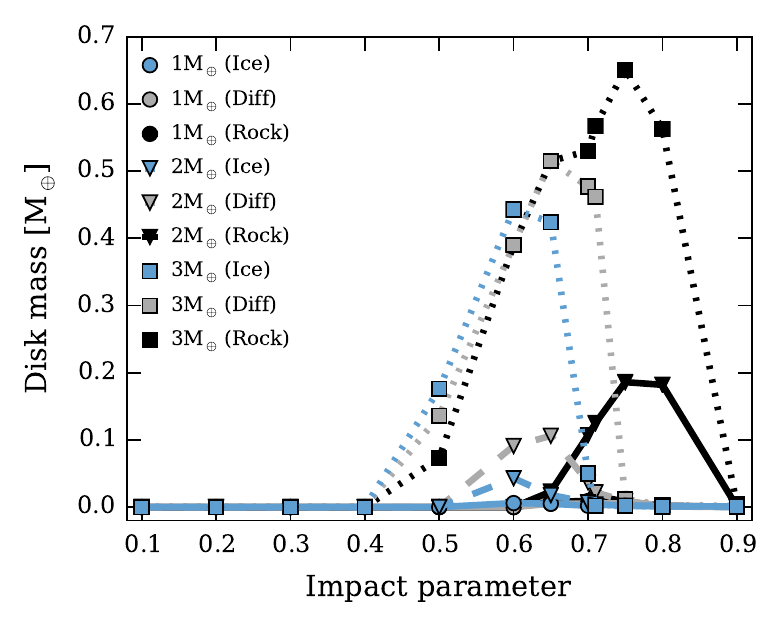}
	
	\caption{\textbf{The mass of the proto-Satellite disk versus impact parameter.} Shown are results for different impactor masses and compositions. Collisions with small impact angles ($b<0.5$) result in an extended, hot atmosphere instead of a disk because the orbiting particle's angular momentum is too small. Grazing impacts do not produce a significant disk because the impactor survives the collision and escapes the system. The impactor's mass and composition clearly affect the disk's mass: the more massive and denser the projectile is, the more material is ejected into the disk.}
	\label{fig:uranus1_diskmass}
\end{figure}

We identify the circumplanetary disk around Uranus as described in Section \ref{subsection:analysis} assuming Uranus' mean density is 1.27~g cm$^{-3}$.
The disks inferred from our simulations have masses ranging from 0.001 to 0.6~\ME
and some of them extend beyond 100~\RE.
Figure \ref{fig:uranus1_diskmass} shows the disk's mass versus the impact parameter for the same impactor mass and composition as in Figures~\ref{fig:uranus1_mass} and ~\ref{fig:uranus1_spin}. We find that disks cannot form for impact parameters $b < 0.4-0.5$ because in these cases the orbiting particles do not have enough angular momentum, and instead they form a spherical envelope/atmosphere. 
Also grazing impacts with $b > 0.8$ do not lead to disk formation because the impactor survives the collision and escapes the planet.
\par

\begin{figure}
	\centering
	\includegraphics{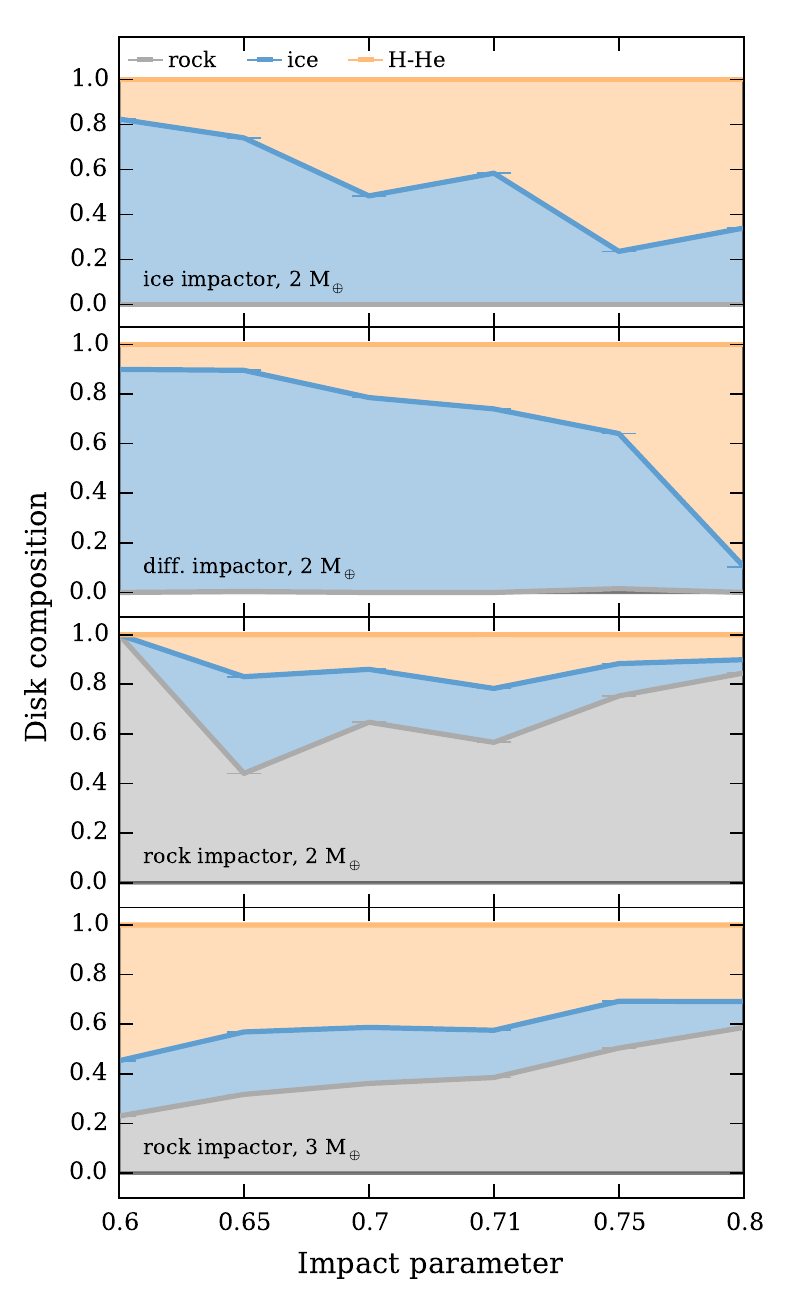}

	\caption{
	\textbf{The composition of the proto-satellite disk.} Shown are the results for two impactor masses, compositions and impact parameters (\vinf=5~km s$^{-1}$, $N=10^5$ particles).
	In all cases it is found that H-He from proto-Uranus' atmosphere, and water from its inner envelope are incorporated into the disk. In some cases differentiated impactors deposit (relative to the disk's mass) more water than icy impactors. The disk composition appears to be insensitive to the impactor mass for 2~\ME and 3~\ME icy and differentiated impactors.
	In order to transfer rock to the disk the impactor must either be differentiated or rocky because the material that originates from proto-Uranus is either from the mantle (ice) or atmosphere (H-He). However, only pure rock impactors can produce disks that are substantially enriched in rock, as is required to explain the composition of Uranus' major satellites. This could change if the differentiated impactor's
	ice-to-rock 
	ratio is varied.
    It may seem counter-intuitive that the relative rock enrichment of the disk is {\em lower} for the more massive rocky impactor. However, the total rock mass deposited in the disk is still larger in this case.
	}
	\label{fig:uranus1_disk_composition}
\end{figure}

We find that the disk's mass increases with increasing impactor mass due to the higher initial angular momentum and kinetic energy of the collision.
Another factor that influences the disk's mass is the assumed impactor's composition: rocky impactors result in more massive disks than icy or differentiated bodies.
Since 10\% - 70\% of the disk's mass originates from the impactor, the impactor's composition substantially affects the inferred disk's composition as shown on Figure~\ref{fig:uranus1_disk_composition} (or Figure~\ref{fig:uranus1_disk_composition_solid} for the heavy-element composition only). None of the collisions with ice impactors result in deposition of rocky material into the disk. This is because the disk material is derived either from the impactor or from the target's ice mantle / H-He atmosphere.
We also observe that a significant fraction of H-He from the target's atmosphere can be incorporated into the disk due to a collision.
However, the forming satellites are not massive enough to accrete a H-He gas envelope from the disk. As a result, the disk's H-He is likely to either be reaccreted by Uranus and/or be lost.
\par

Forming a proto-satellite disk is the first step.  
Then, one must ensure that the disk consist of enough mass in heavy elements (i.e., rock and ice) and is sufficiently extended in order to explain the formation of  Uranus’ regular  satellites \citep{Morbidelli2012}. 
For most of the disks obtained in our simulations 90\% of the mass is contained within 20 to 90~\RE, i.e. there is often less than 1\% of the mass beyond the orbit of Oberon, Uranus' most outer regular satellite. Since most disks are very massive (> 0.1~\ME) only a tiny fraction of the total disk mass (corresponding to less than 10 particles in the $10^5$ particle simulations) is required to form Oberon.
Finally, the disk should have the appropriate composition. 
The regular moons of Uranus are composed of about 50\% rock and 50\% ice which means that the satellite disk should consist of enough rocky material.
We thus define a potential Uranus proto-satellite disk as a disk that: (i) contains at least the total mass of Uranus’ regular satellites $M_S = 1.5 \times 10^{-3}$~\ME in rock or ice, (ii) extends beyond 93~\RE which is Oberon's distance and contains at least its mass ($5 \times 10^{-4}$~\ME) in rock and ice beyond this distance, and (iii) has a minimum rock mass of half the total satellites mass. 
According to this definition, ten of the simulations presented in Figure~\ref{fig:uranus1_diskmass} (e.g., AU2g8-11 with $b=0.7-0.8$, AU3g6-11 with $b=0.6-0.8$) lead to the formation of potential Uranus' proto-satellite disks.
It is found that none of the differentiated impactors deposit enough rocky material in the disk. This, however, could change when considering lower ice-to-rock ratios (i.e., a larger rock fraction) for the differentiated impactor.

\subsection{Hit-and-run collisions}
\label{subsection:hit-and-run-collision}
We also investigate HRC on proto-Uranus.  HRC 
are characterised by a large initial amount of angular momentum and small mass exchange between the bodies.
Such an impact can explain Uranus' tilt and because little mass is exchanged in the collision, also the small mass difference between Uranus and Neptune, and the fact that Uranus' interior is more centrally concentrated and possibly non-convective. 
As an extreme case we consider a grazing ($b=0.6-0.7$) collision of Uranus with a twin planet of the same mass and composition (for example an ejected fifth giant planet as suggested by \citealt{Nesvorny2011}).
We vary the velocity at infinity to be between 1.5 to 4~\vesc, resulting in  impact velocities between 30 and 45~km s$^{-1}$, velocities that are significantly larger than Uranus' current orbital velocity of $\sim6$~km s$^{-1}$. 
We find that none of these collisions reproduce Uranus' spin from a non-rotating target, and that the inferred rotation period is always larger than 30 hours.
This is also true when we consider initial rotation periods of $P=20$~h or $P=30$~h because the escaping projectile removes most of the angular momentum from the system.
We therefore conclude that such HRC are unlikely to explain Uranus' observed properties.

%%%%%%%%%%%%%%%%%%%%%%%%%%%%%%%%%%%%%%%%%%%%%%%%
% Neptune
%%%%%%%%%%%%%%%%%%%%%%%%%%%%%%%%%%%%%%%%%%%%%%%%
\section{Neptune}
\label{section:neptune}

%%%%%%%%%%%%%%%%%%%%%%%%%%%%%%%%%%%%%%%%%%%%%%%%
% Fig: Total bound mass after the impact 
%%%%%%%%%%%%%%%%%%%%%%%%%%%%%%%%%%%%%%%%%%%%%%%%
\begin{figure}
	\centering
	\includegraphics{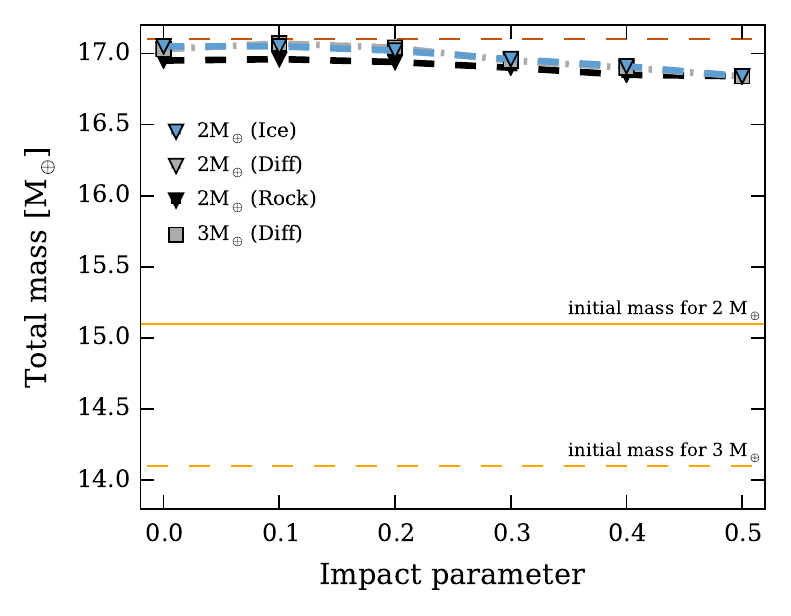}

	\caption{\textbf{Post-impact total bound mass of the Neptunian system (planet+envelope+disk) for different impactor masses and compositions.}
    The total colliding mass is set to Neptune's observed value (17.1~\ME, dashed red line). The different symbols represent different impactor masses (triangle: 2~\ME, square: 3~\ME) and the colours correspond to the impactor's composition (blue: ice, grey: differentiated, black: rock).
	The orange lines show the planet's initial mass which depends on the impactor's mass (dashed: 2~\ME and dotted: 3~\ME).
	In all cases the target is represented with $10^5$ particles (see Section \ref{subsection:ic} for details).
	Most of the impactor is accreted for collisions with $b < 0.6$ (see Section~\ref{section:neptune} for details).}
	\label{fig:neptune1_mass}
\end{figure}

%%%%%%%%%%%%%%%%%%%%%%%%%%%%%%%%%%%%%%%%%%%%%%%%
% Fig: head-on and grazing impact (Mi=2ME, 5M)
%%%%%%%%%%%%%%%%%%%%%%%%%%%%%%%%%%%%%%%%%%%%%%%%
\begin{figure}
	\centering
	\includegraphics[keepaspectratio,width=0.46\textwidth]{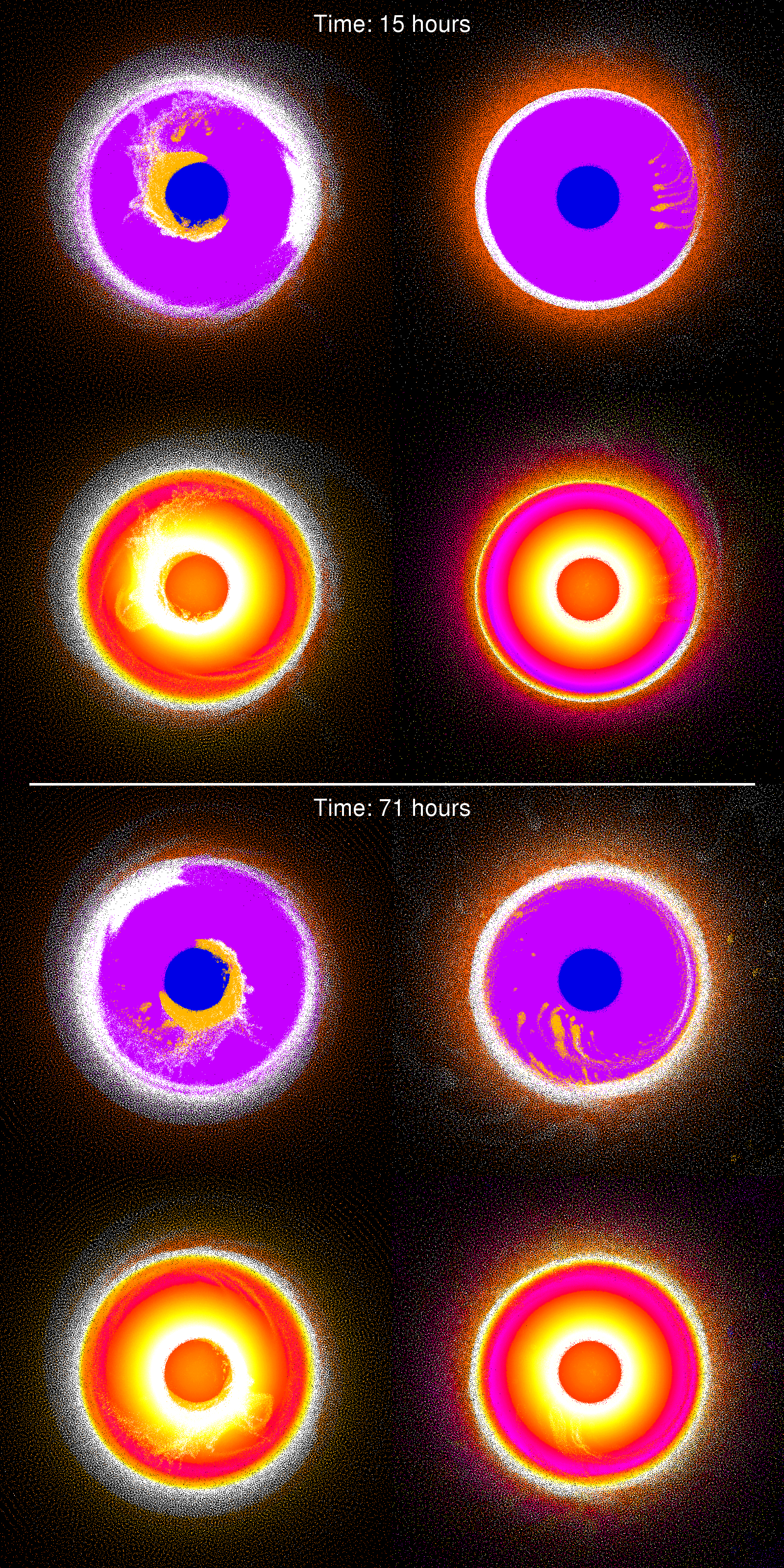}

	\caption{\textbf{The planet's interior after a head-on (left) and a grazing (right) collisions}.
	Shown are the results for a giant impact on Neptune (differentiated 2~\ME impactor, $N=5 \times 10^6$ particles, \vinf=5~km s$^{-1}$) for $b=0.2$ (head-on, left) and $b=0.7$ (grazing, right) 15~h (top panel) and 71~h (bottom panel) after the impact. 
	The size of an individual snapshot is 8 \RE $\times$ 8 \RE $\times$ 1 \RE. The top panels shows the origin of the material (target core: blue, mantle: violet, atmosphere: orange and impactor core: yellow, mantle: white).
	The bottom figures show the
	internal energy of the particles between 0 erg g$^{-1}$ (black) and $10^{12}$ erg g$^{-1}$ (white).
	For the head-on collision the projectile's core and part of its mantle penetrates deeply into the target. The atmosphere but also the planet's interior are substantially heated.
	In case of the grazing collision, during the the initial impact (top panel) the projectile only interacts with the target's atmosphere and upper mantle so it survives the first encounter.
	Much less material and energy is deposited in the planet and most of it remains in the atmosphere and upper mantle. 
	The impactor remains bound to the planet and re-impacts two days later.
	This second collision is more head-on but since the projectile's core is eroded during the tidal encounter it can not impact the planet's core and the rock is distributed in the mantle. In both cases $\sim$10\% of the original H-He is ejected because of the impact.
	}
	\label{fig:neptune_hvsg_t8.7_40}
\end{figure}

For Neptune we focus on head-on collisions ($b=0.0-0.5$) that result in accretion of the impactor (Figure \ref{fig:neptune1_mass}). 
Such collisions could explain the higher mass of Neptune in comparison to Uranus, and result in a higher moment of inertia value for Neptune. 
This is because such impacts are expected to deposit sufficient amounts of energy and mass in the planetary deep interior that could lead to mixing and to a temperature gradient that is closer to an adiabatic one resulting in a convective interior (e.g., \citealt{Podolak2012}). 

%%%%%%%%%%%%%%%%%%%%%%%%%%%%%%%%%%%%%%%%%%%%%%%%
% Fig: Target after the impact (b=0.1, Mi=2ME)
%%%%%%%%%%%%%%%%%%%%%%%%%%%%%%%%%%%%%%%%%%%%%%%%
\begin{figure*}
	\centering
	\includegraphics[keepaspectratio,width=\textwidth]{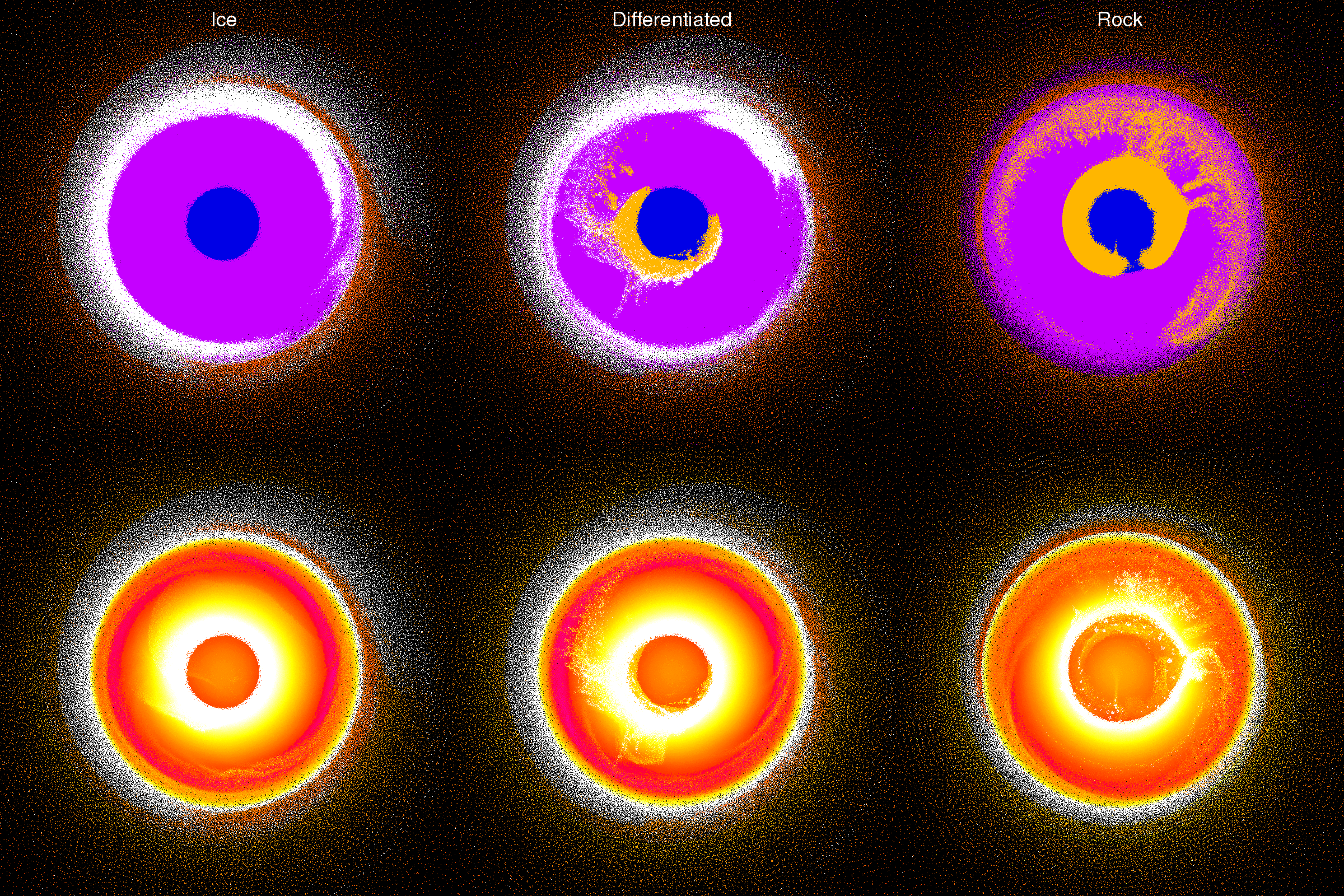}

	\caption{\textbf{The distribution of the impactor's material in a head-on collision for different impactor composition.} The results correspond to Uranus resolved with $5 \times 10^6$ particles after colliding with a 2~\ME impactor at $b$=0.2 for various impactor  composition (from left to right: ice, differentiated, and rock).
	The box size is 8~\RE $\times$ 8~\RE $\times$ 1~\RE and the colors correspond to the origin of the material (see Figure~\ref{fig:neptune_hvsg_t8.7_40}).
	A pure-ice impactor is stopped in the target's upper mantle, while 
	a pure-rock impactor penetrates deep into the planetary interior depositing most of its mass on above the core. During the passage trough the inner envelope, the projectile is partially eroded and leaves rock in the planet's outer regions. 
	In case of a differentiated impactor most of the projectile's mantle remains in the planet's upper envelope while the projectile's core penetrates deep into the planet and impacts the core. In this case both water and rock from the impactor are deposited within planetary interior.
    }
	\label{fig:neptune_imp_comp}
\end{figure*}

The outcome of an impact on Neptune is very similar to a head-on collision on Uranus (see Figure \ref{fig:neptune_hvsg_t8.7_40} for an example). 
The projectile easily penetrates the gaseous envelope and hits the target's mantle. 
The exact outcome depends on the impactor's composition as shown in Figure~\ref{fig:neptune_imp_comp}. A pure-ice impactor deposits all of its mass in the planetary upper mantle for all impactor masses and resolutions considered.
There it forms a layer of shocked, hot material that can have a different composition from the surrounding mantle material.
Larger impact parameters lead to larger areas that are covered by this hot material. 

%%%%%%%%%%%%%%%%%%%%%%%%%%%%%%%%%%%%%%%%%%%%%%%%
% Fig: Where is the energy deposited
%%%%%%%%%%%%%%%%%%%%%%%%%%%%%%%%%%%%%%%%%%%%%%%%
\begin{figure}
	\centering
	\includegraphics{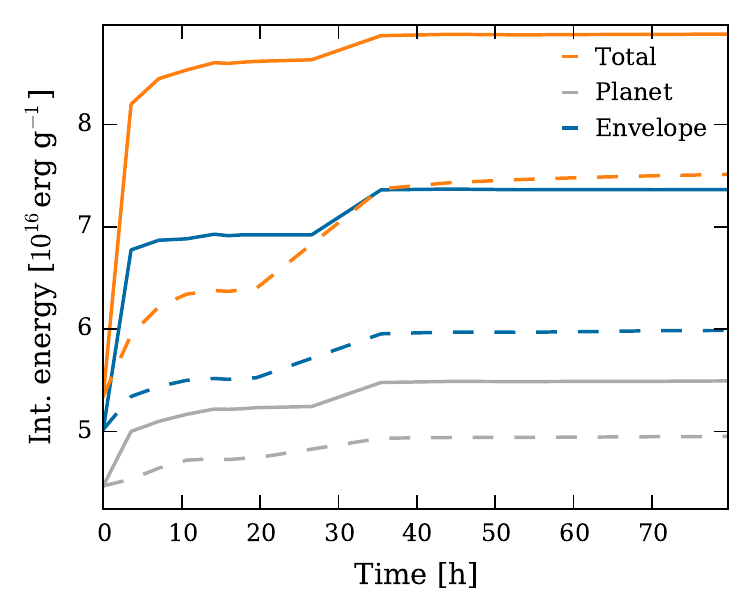}

	\caption{\textbf{The target's internal energy distribution for a head-on versus a grazing collision.} We show the targets' internal energy after colliding with a 2~\ME differentiated body at $b=0.2$ (solid lines) and $b=0.7$ (dashed lines) with \vinf=5~km~s$^{-1}$ using $10^5$ particles. In both cases the envelope (blue) absorbs a significant fraction of the total energy (orange) deposited in the collision.
	However, the head-on collision deposits more energy in total and also more energy in the planetary  interior (grey). Since the impactor's remnant survives the first encounter with the planet in the grazing collision, the energy is deposited in two steps: during the initial impact and when the impactor's remnant collides a second with the planet at time $t$ 35~h.
	}
	\label{fig:neptune_profile_int_energy}
\end{figure}

%%%%%%%%%%%%%%%%%%%%%%%%%%%%%%%%%%%%%%%%%%%%%%%%
% Fig: Rotation period after the impact 
%%%%%%%%%%%%%%%%%%%%%%%%%%%%%%%%%%%%%%%%%%%%%%%%
\begin{figure}
	\centering
	\includegraphics{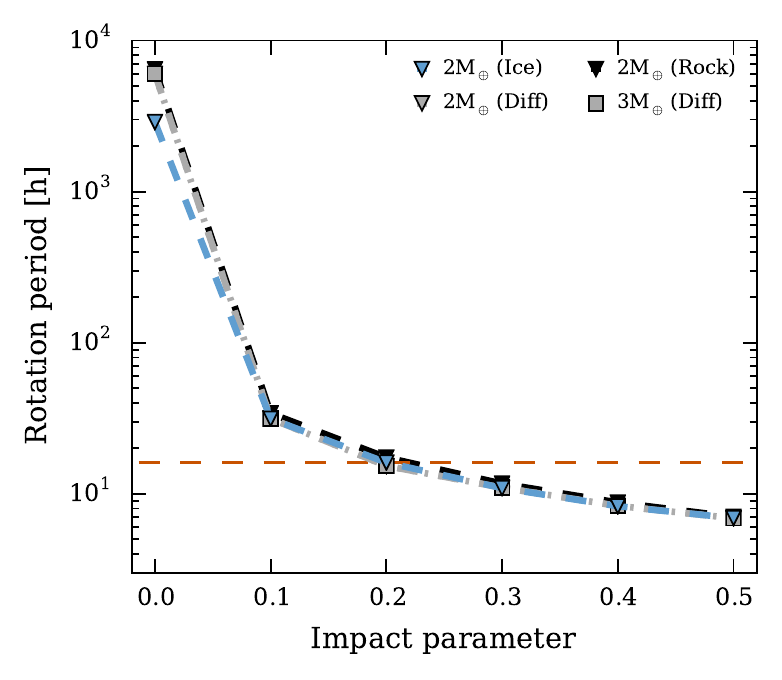}

	\caption{\textbf{Neptune's post-impact rotation period for different impactor masses and compositions.} Neptune's current rotation period of (16.11~h) is shown with the dashed 
	red line. The different symbols and colours correspond to different impactors masses and compositions as indicated in the legend. Note that the vertical axis is in log-scale. 
    The initial conditions are set as described in Section \ref{subsection:ic}, with a non-rotating proto-Neptune (resolved with $10^5$ particles) prior to the collision. Neptune's rotation period can be reproduced in collisions with $b>0$. For a strictly head-on collision ($b=0$), the period is very large as almost no angular momentum is transferred to the target.
	}
	\label{fig:neptune1_spin}
\end{figure}

Such a collision adds mass and energy to the planet (Figure \ref{fig:neptune_profile_int_energy}), as well as angular momentum (see Figure \ref{fig:neptune1_spin}).
As the planet cools down and relaxes from the post-impact state, material and energy could be redistributed, due to convective mixing. It is therefore desirable to model the post-impact long-term evolution of the planets and investigate how impacts can affect the density distribution within the planets, and possibly, explain the inferred differences in the MOI values of Uranus and Neptune (e.g., \citealt{Podolak2012}).  

We also observe that the treatment of the interfaces (see Section~\ref{subsection:densitycorrection}) affects the detailed way in which the impactor's water ice is distributed in the upper mantle. However, the general behaviour agrees with standard SPH, even for head-on collisions ($b=0.2$); icy impactor material is uniformly distributed above the planet's mantle. 
Rocky projectiles on the other hand hit the core, depositing mass and energy deep inside the planet. On its way in, the projectile loses mass as it passes through Neptune's mantle, enriching the icy mantle with rocky material from the impactor.  
The exact mass of rocky material that is deposited into the icy mantle depends on the impact angle, and the resolution. The larger the impact parameter, the longer the projectile interacts with the ice layer,  decreasing the ice-to-rock ratio in the upper mantle.

%%%%%%%%%%%%%%%%%%%%%%%%%%%%%%%%%%%%%%%%%%%%%%%%
% Fig: Interior for different res (b=0.2)
%%%%%%%%%%%%%%%%%%%%%%%%%%%%%%%%%%%%%%%%%%%%%%%%
\begin{figure*}
	\centering
	\includegraphics[keepaspectratio,width=\textwidth]{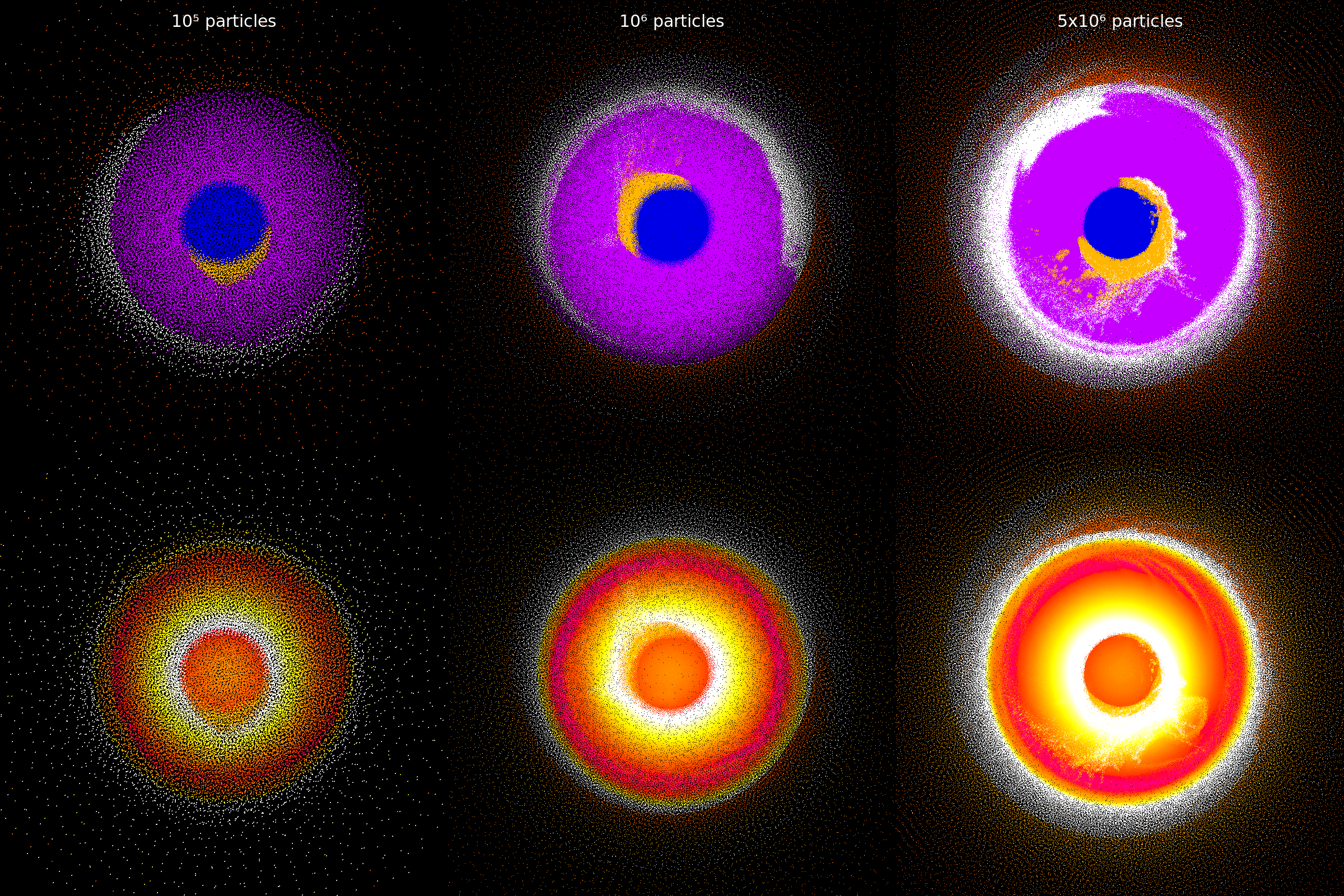}

	\caption{\textbf{The distribution of the impactor's material within the planet using different resolutions (from left to right: $10^5$, $10^6$ and $5 \times 10^6$ particles for the target)}. 
	The distributions correspond to 70.7~h 
	after the head-on collision presented in Figure \ref{fig:neptune_hvsg_t8.7_40}.
	If the target is resolved with $10^5$ particles (left) no materials from the impactor is mixed into the planet's mantle. 
	Increasing the resolution to $10^6$ particles (middle), the stripping of the impactor when it passes through the planet's mantle is resolved. When we use $5 \times 10^6$ particles (right) ice and rock from the impactor are clearly mixed into the planet's mantle affecting its composition and thermal profile (as shown in Figure~\ref{fig:neptune_hvsg_t8.7_40}). 
	}
	\label{fig:neptune_mat_100k_vs_5M}
\end{figure*}

%%%%%%%%%%%%%%%%%%%%%%%%%%%%%%%%%%%%%%%%%%%%%%%%
% Fig: Enrichment of the mantle
%%%%%%%%%%%%%%%%%%%%%%%%%%%%%%%%%%%%%%%%%%%%%%%%
\begin{figure}
	\centering
	\includegraphics{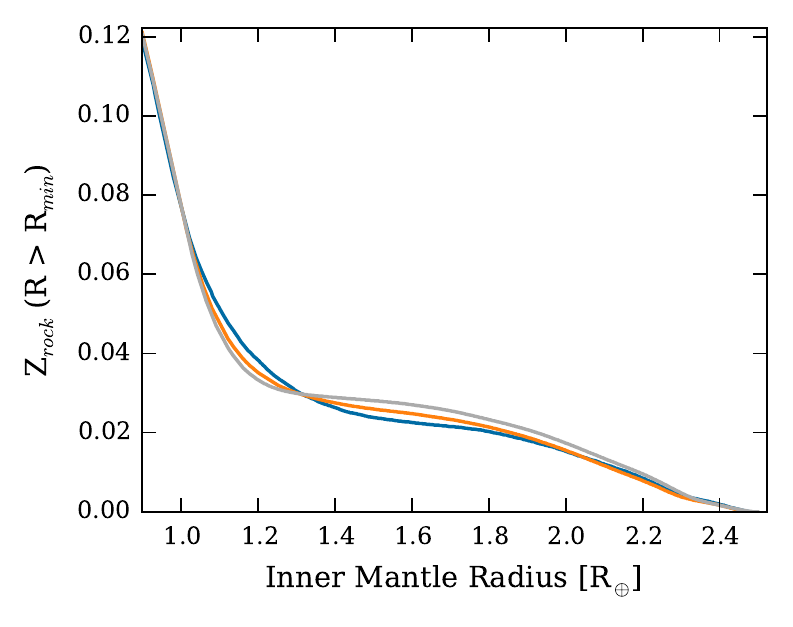}

	\caption{\textbf{An example of the enrichment of the (initially pure-) water layer in a head-on collision for different resolutions}.
	Shown is the enrichment m$_{rock}$ of the planet's ice layer in impactor's rock after a head-on collision with a rocky 2~\ME impactor ($b=0.2$, \vinf=5~km s$^{-1}$) for different resolutions ($10^5$: blue, $10^6$: orange and $5 \times 10^6$: grey).
	Since the transition from the planet's core and mantle after the collision is not well-defined the inferred enrichment can vary strongly for small radii. An increased inner radius for the water layer lead to more similar values for all resolutions. However, higher resolutions result in a higher mantle enrichment because the projectile's erosion is better resolved.
	}
	\label{fig:neptune_mantle_enrichment}
\end{figure}

While the inferred total rock mass deposited in the planet's mantle for a given impact conditions (impactor mass, impact parameter and velocity) agrees for all resolutions, it is found that simulations with $10^5$ particles do not resolve the location where the rock is deposited, and most of it remains near the planet's surface.
Increasing the resolution provides a clearer picture as the impactor's erosion and the deposition of its material in Neptune's mantle is well-resolved (see Figure \ref{fig:neptune_mat_100k_vs_5M} for an example of how the resolution affects the material distribution in case of a differentiated impactor).
For simulations with $5 \times 10^6$ particles, it is found that a rocky impactor leads to the formation of a thick blanket of enrichment in the planet's upper mantle. 
This is also reflected in the inferred enrichment of the planet's mantle.
For example, in case of a head-on collision ($b=0.2$) we obtain a rock mass fraction $m_{rock}$ of $\sim$ 2\% in a $10^5$ particle simulation.
When $5 \times 10^6$ particles are used we obtain $m_{rock} = 3\%$ which is 50\% higher than in the lower resolution case (Figure \ref{fig:neptune_mantle_enrichment}).
While the enrichment values vary for different impact conditions such as the impact parameter, impactor mass and composition, the general behaviour is expected to remain. Increased resolution reveals more details on the material's deposition and leads to higher enrichment. Our results demonstrate that GIs can increase the rock mass fraction in the ice giants. 
Again, for an increasing impact parameter, more rock is mixed within the mantle and more of the planet's upper mantle is  covered by this blanket of enrichment (see Figure~\ref{fig:neptune_imp_comp}).

If the impactor is differentiated an intermediate scenario occurs. Upon hitting the target's mantle the projectile breaks apart, the ice remains in the upper mantle while the core can penetrate deeper.
Thus it seems that in order to affect Neptune's interior the projectile should preferably be composed of (at least some) refractory material. For all the cases we consider, it is found that larger impact velocities and smaller impact angles lead to a more significant effect on Neptune's deep interior. It is also found that the H-He atmosphere absorbs a substantial part of the impact energy.

Also here, the resolution of the simulation plays an important role. For the higher resolution simulations ($10^6$ or more particles), the impactor is more eroded, enriching the icy shell with rocky material.
We also observe that more ice from the impactor's mantle is mixed in Neptune's deep interior using higher resolutions (Figure~\ref{fig:neptune_mat_100k_vs_5M}) which never occurs for a pure-ice projectile. 

As discussed in Section~\ref{subsection:rotation_period} for Uranus a 2~\ME impactor can induce rotation periods below $17$~h in head-on collisions. For Neptune the general trend is found to be very similar for a given impactor mass and composition. Since for Neptune the preferred collisions are ones with $b \sim 0.2$, the inferred rotation periods are of the order of 15 hrs, which is consistent with the measured Voyager period. Due to the slightly higher angular momentum of the collision for the case of Neptune, its rotation period tends to be $\sim 5\%$ higher than Uranus. This is consistent with the modified rotation periods of the planets as suggested by \citep{Helled2010}. 

%%%%%%%%%%%%%%%%%%%%%%%%%%%%%%%%%%%%%%%%%%%%%%%%
% Discussion and conclusion
%%%%%%%%%%%%%%%%%%%%%%%%%%%%%%%%%%%%%%%%%%%%%%%%
\section{Discussion}
\label{section:discussion}
We simulate giant impacts on Uranus and Neptune accounting for various impact angles and velocities, impactor mass and composition and numerical parameters (e.g., resolution, viscosity limiter and interface correction). 
We investigate whether Uranus' tilt and the observed difference in thermal flux between Uranus and Neptune can be explained by such impacts. 
For Neptune we investigate whether a head-on collision can deposit enough mass and energy in its deep interior leading to a hotter and less centrally concentrated interior in comparison to Uranus. This has the potential to explain the differences in the MOI values and heat fluxes of the ice giants. Interestingly, such an impact also leads to a small increase in Neptune's mass which could explain the differences in mass between the two planets. While this is very speculative, it clearly reflects the potential influence of giant impacts on the planetary characteristics. 
Head-on collisions also do not produce a proto-satellite disk, consistent with Neptune's irregular major satellites.
The initial spin of both planets are unknown and the impact conditions that lead to Uranus' tilt of 97$^{\circ}$ depend somewhat on the target's pre-impact spin. In this work we consider only non-rotating targets with the exception of extreme HRC, where proto-Uranus collides with an ejected twin planet of the same mass. In this case we assigned proto-Uranus an initial rotation period of $P=30$~h and $P=20$~h.

For Uranus we find that its rotation period of 17.24~h can be produced in most of our simulations.
The impactor's mass and composition clearly affect the rotation period: more massive bodies have a larger initial angular momentum and thus induce a smaller rotation period. 
In addition, low-density (i.e. icy) impactors contribute more angular momentum for small impact parameters because most of the mass remains in the outer mantle of Uranus. Conversely, for larger impact parameters icy impactors also enter the hit-and-run regime at lower impact parameter than rocky ones and therefore are less efficient in increasing the planet's angular momentum.

While our inferred  trend agrees well with previous work (\citealt{Slattery1992} and \citealt{Kegerreis2018}), we find that in most cases also a 1~\ME impactor can reproduce Uranus' rotation. These bodies were excluded as candidates to explain Uranus' tilt in earlier investigations by S92 and K2018 because they could not deposit enough angular momentum in the planet.
In both studies the total angular momentum of the collision was used to parametrise the collision, while in our simulation the initial conditions are described in terms of the impact parameter.
This complicates a direct comparison of the results.
Since the impact velocity depends strongly on the systems escape velocity, differences in proto-Uranus radius can affect the impact velocity, and thus the initial angular momentum of the collision.
Another potential explanation for this difference is the EOS used to model the various materials, especially the H-He envelope.
In order to investigate the sensitivity of the results to the used EOS for H-He, 
we consider an extreme case of a solid initial proto-Uranus with a rock core and an ice mantle (Appendix \ref{appendix:two_component_models}). We find that the inferred rotation periods are very similar to the one obtained for the  three component model (as discussed in Section~\ref{subsection:models}) of the same mass.
\par

We also find that Uranus' rotation period depends on the simulation's resolution: while the rotation period converges quickly after the impact and remains constant in the low resolution simulations we observe transport of angular momentum from the planet to the envelope caused by artificial viscosity in the higher resolution simulations. Using a viscosity limiter or further increasing the resolution reduces the decay of the rotation period over time.

It should be noted that other explanations for the properties of Uranus and Neptune have been proposed. For example, \citet{Boue2010} showed that Uranus' tilt can be the result of interactions between the planet and an additional massive satellite during migration in the protoplanetary disk. Similarly Neptune's obliquity of $29.5 \degree$ has also been proposed to be excited during its migration \citep{Parisi2011}. 
Also the origin of Uranus' proto-satellite disk also does not have to be due to a collision. 
Alternatively, the planet could have accreted a circumplanetary disk during its formation \citep{Szulagyi2018} before having the protoplanetary disk tilted due to spin orbit interaction in a suite of
collisions involving less massive impactors or lower impact velocities than investigated here
\citep{Morbidelli2012}.
However these alternative scenarios do not solve the internal structure dichotomy. In addition, the relatively large obliquity of both planets ($\gtrsim 30 \degree$) is quite consistent with having experienced at least one violent collision after their formation.
\par 

While a GI is not the only possible explanation, 
the retrograde rotation of Uranus' five major satellites can  be explained if the same collision that tilted the planet also led to the formation of a circumplanetary disk.
Many of our simulations lead to the formation of massive and extended disks. However, most of them have less than the minimum amount of rocky material needed to form the regular satellites with a 50\% rock composition. Since the disk's material either originates from the impactor or the target's upper layers, we suggest that an impact of a rock-rich object is more likely. 
A differentiated impactor can also deposit rocky material in the disk due to the tidal disruption of the core but not enough to form all satellites.
Only a rocky impactor produces disks that satisfy all constraints and we find several good candidates amongst our simulations. 
However, differentiated impactors with 
lower ice-to-rock 
ratios than we chose could again produce the desired satellite disk properties. 
In addition, if the impactor is composed of a mixture of rock and ice and is undifferentiated, or if the rock is mixed into Uranus' outer envelope the resulting disk can be further enriched in rock.

High resolution simulations show that a pure-rock impactor can substantially enrich Uranus' mantle with rock in a grazing collision.
In all cases the disk contains more than 10\% (by mass) H-He from Uranus' atmosphere. These findings could have consequences for the internal structure and thermal evolution of Uranus as well as  the formation of its satellites. 
In K2018 the disk is defined as all the orbiting material outside the Roche limit because close to the planet tidal forces prevent satellite formation.
Because in our simulations the disks are rather massive and only $\sim$10\% of the mass is inside of the Roche limit including the above constraint to our definition of a proto-satellite disk does not affect our conclusions. In addition, material closer to the planet can be used for satellite formation at later stages due to viscous spreading of the disc (\citealt{Salmon2012}, \citealt{Crida2012}). 
Another open question is how much of the orbiting material can be accreted and form satellites. This depends on several factors, e.g., planet mass and the physical conditions in the disk. Depending on how effective material is accreted and how much material is ejected or reaccreted by the planet some of the proto-satellite disk candidates found in our simulations can be excluded.

For Neptune we find that head-on collisions deposit impactor material and energy deep in its interior. Independent of the impactor's mass or resolution, ice usually remains in the upper mantle and atmosphere (see Figure \ref{fig:neptune_imp_comp}) so icy projectiles are unlikely to affect the internal structure significantly.
On the other hand, rocky or differentiated impactors penetrate into the deep interior of the planet.
In such collisions, the impactor's rocky material (and in case of a differentiated projectile also some ice) is deposited deep into Neptune's mantle, and the mass and energy are deposited near the core. It is also found that large impact parameters, e.g., $b \sim 0.5$, lead to more mixing of the impactor's material into the planetary mantle and to a more homogeneous internal structure. 

We also find that the simulation's resolution plays a key role when investigating the effect of GI on the planetary interior. 
First, for higher resolutions ($10^6$ particles or more) the rotation period does not settle down to a single value due to an artificial angular momentum transport from the planet to the outer envelope. Resolution also plays a role when studying the disk and the planet's post-impact composition, affecting the outcome of the simulation in terms of mixing. The impactor's erosion in the planet's mantle and thus the heavy-element enrichment (rock, water) can only be resolved when using $>10^6$ particles. In addition, higher resolution leads to more mixing of the impactor's rock and ice material in the planet's mantle. Head-on collisions of differentiated impactors deposit the impactor's ice near the core when using $5 \times 10^6$ particles.
Finally, a resolution of $>10^6$ particles is required to observe the tidal disruption of a differentiated impactor's core in grazing collisions. This has profound implications for the distribution of the impactor's rock in the post-impact planet (Figure \ref{fig:neptune_hvsg_t8.7_40}). If the impactor's core is not tidally eroded after the first collision, it merges with the planet's core during the second collision. Otherwise, small rocky clumps fall back onto the planet and deposit the rocky material in the planet and/or  disk.

Clearly, giant impacts can significantly affect the planetary internal structure. However, our simulations are limited to several
 days 
after the impact. The next required step is to use the output of the impact simulations (energy, composition) and model the long-term thermal evolution of the planets. This can reveal whether GI can indeed explain the differences in heat flux and internal structure (e.g. MOI) between the two planets as implied by \citet{Podolak2012}. 
This is particularly important for Neptune since it can allow an investigation of whether the energy and mass associated with the impact can lead to convective mixing and a more homogeneous interior.  
Also for Uranus it is important to investigate whether the inferred layered structure 
can persist during the planet's long-term evolution as grazing impacts can induce differential rotation in the planet's outer region and therefore promote mixing (e.g., \citealt{Nakajima2015}).
\par

\section{Conclusions:}
Our main conclusions can be summarised as follows:
\begin{itemize}
    \item Giant impacts can explain the observed differences between Uranus and Neptune. 
    \item Giant impacts on Uranus and Neptune can substantially alter their rotation axis and internal structure.
    \item Uranus' current rotation period can be produced in most of our simulations.
    \item A giant impact on Uranus can lead to the formation of an extended disk providing enough material for the formation of its regular satellites after the collision.
    \item Hit-and-run collisions cannot alter the target's rotation axis and do not lead to the formation of a proto-satellite disk even when a rotating target is assumed.
    \item Head-on collisions for Neptune result in accretion of more mass and energy, and substantially affect the planet's interior. 
    \item Our simulations favour impactors that are substantially enriched in rock in order to explain the dichotomy between Uranus and Neptune.
\end{itemize}

Our work suggests that Uranus and Neptune could have had similar properties (masses, internal structures) shortly after their formation and that the observed differences between the planets (tilt, satellite system, flux) are caused by giant impacts with different conditions. Given the large number of impacts during the early days of the solar system this scenario is appealing. It is also interesting to note that giant impacts are thought to play an important role
in explaining the characteristics of the inner planets such as Mercury's high iron-to-rock ratio (e.g., \citealt{Benz2007,Asphaug2014a,Chau2018}), the Earth's Moon (e.g., \citealt{Canup2001a,Deng2019a}), and Jupiter's dilluted core (e.g., \citealt{Liu2019}).
This emphasises the role of giant impacts for our understanding of planetary objects. 

Clearly, there is still much more work to be done, and this study only represents the beginning of a long-term investigation of the role of giant impacts in understanding Uranus and Neptune. Future investigations should include: (i) simulations of the post-impact thermal evolution of the planets. (ii) use constraints from N-body simulations to better determine Uranus' pre-impact rotation and the likelihood of the various impact conditions. 
(iii) use more realistic equations of state for the assumed materials. (iv) consider a larger range of compositions and internal structures for the targets and impactors. 
(v) higher resolution simulations in order to better resolve Neptune's interior and Uranus' proto-satellite disk.

Uranus and Neptune represent a unique class of planets in the solar system, and yet, they are poorly understood. 
The upcoming years are expected to include new studies about these planets given the increasing interests of both ESA and NASA to send dedicated spacecraft to these planets, and the fact that a large fraction of the discovered exoplanets have similar masses/sizes to those of Uranus and Neptune. We therefore hope that we are at the beginning of a new era in ice giant exploration. 

\section*{Acknowledgements}
C.R and J.S.~acknowledge support from SNSF Grant in ``Computational Astrophysics'' (200020\_162930/1). 
R.H.~acknowledges support from SNSF grant 200021\_169054.
We also thank D.~Stevenson and M.~Podolak for valuable discussions and suggestions.
We also thank the anonymous referee for valuable suggestions and comments that helped to improve the paper.
The simulations were performed using the UZH HPC allocation on the Piz Daint supercomputer at the Swiss National Supercomputing Centre (CSCS). This work has been carried out within the framework of the National Centre of Competence in Research PlanetS, supported by the Swiss National Foundation.

%%%%%%%%%%%%%%%%%%%%%%%%%%%%%%%%%%%%%%%%%%%%%%%%
% The appendix. Here I moved all the stuff that
% might be interesting but does not go into the
% main part of the paper.
%%%%%%%%%%%%%%%%%%%%%%%%%%%%%%%%%%%%%%%%%%%%%%%%
\begin{appendix}
%%%%%%%%%%%%%%%%%%%%%%%%%%%%%%%%%%%%%%%%%%%%%%%%
% Two component models.
%%%%%%%%%%%%%%%%%%%%%%%%%%%%%%%%%%%%%%%%%%%%%%%%
\section{Two component models}
\label{appendix:two_component_models}
In order to investigate the sensitivity of the results on the used EOS for the H-He atmosphere, we 
perform several impact simulations using a two component target that consists of a 10\% by mass rocky core and a 90\% ice mantle.
Obviously, solid ice is a poor choice when attempting to model an enriched H-He atmosphere but it provides an upper limit on the atmosphere's interaction with the projectile. The models and initial conditions are generated as described in Section~\ref{subsection:ic} and the target is resolved with $10^5$ particles.

\begin{figure}
	\centering
	\includegraphics{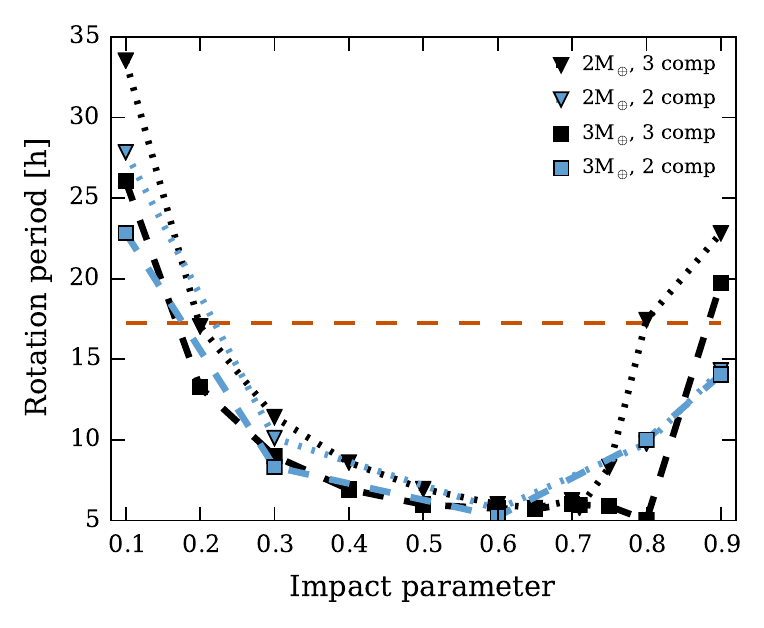}

	\caption{
	\textbf{Uranus' post-impact rotation period for three-component vs. two-component initial models.}
	We show the rotation period for a two-component (blue,  see Appendix~\ref{appendix:two_component_models} for details) and a three-component (black) model of proto-Uranus.
	Both simulations are resolved with $10^5$ particles. 
	Uranus’ current rotation period of (17.24 h) is shown with a dashed orange line. 
	The impactor is assumed to be rocky, and we consider masses of both  2~\ME (triangle) and 3~\ME (square). Overall, the rotation periods are in good agreement. The two-component models are found to have slightly faster rotation than the three-component ones, with the difference being most pronounced for very head-on and grazing impacts. 
	}
	\label{fig:uranus_spin_23comp}
\end{figure}

Generally, the results of the simulations are found to be similar.
Rock from the impactor also impacts with the target's core, except when the projectile is tidally disrupted. The ice material remains mostly in the outer regions of the inner envelope of the planet. Only for very small and very large impact parameters we observe a difference as the the projectile cannot penetrate as easily to the target's ice mantle  as in the case of a H-He  atmosphere. In that case, a larger fraction of the projectile remains in the upper mantle and  more angular momentum is transferred to the planet, and as a result the inferred rotation period is affected (see Figure~\ref{fig:uranus_spin_23comp}). In addition, grazing collisions ($b > 0.75$) lead to more mergers compared to the three component models. Since the two component models are more compact, the impact velocity is found to be slightly higher (for details see Section~\ref{section:discussion}).

%%%%%%%%%%%%%%%%%%%%%%%%%%%%%%%%%%%%%%%%%%%%%%%%
% Numerical test.
%%%%%%%%%%%%%%%%%%%%%%%%%%%%%%%%%%%%%%%%%%%%%%%%
\section{Numerical tests}
\label{appendix:box_test}
\begin{figure*}
	\centering
	\includegraphics[keepaspectratio,width=\textwidth]{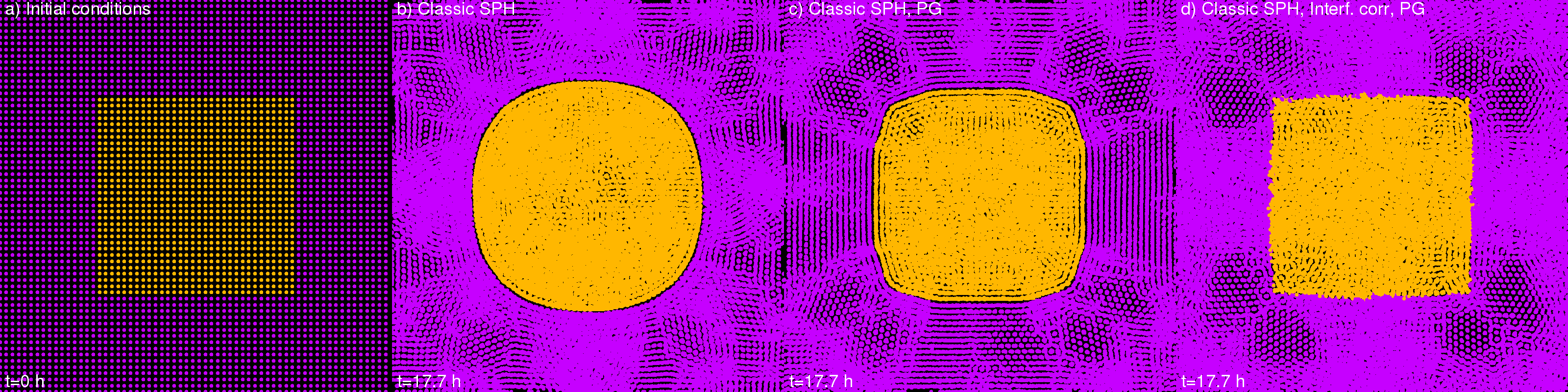}

	\caption{\textbf{A thin slice ($L_x$=0.5 \RE, $L_y$=0.5 \RE, $L_x$=0.001 \RE using periodic boundary conditions) through the results for the box test using iron (yellow) and granite (violet) initially (a) and after 17.7~h in simulation time for the different flavours of SPH used in this work.} As reported in previous work standard SPH (b) suffers from artificial pressure forces at the material interface which acts as a surface tension that quickly causes the iron box to assume spherical shape. Using the geometric density average force (GDF) (c) \citep{Wadsley2017} already reduces this effect but the result is clearly better when an ideal gas EOS is used. Only when the material interface treatment proposed in this paper is combined with GDF (d) the initial box remains stable over the whole simulation (17~h) and the corners are well resolved.}
	\label{fig:boxtest_iron_granite}
\end{figure*}

As mentioned in Section \ref{subsection:densitycorrection}, SPH cannot properly handle contact discontinuities. One popular test to investigate a SPH code's performance in such a situation is the box test \citep{Saitoh2013}.
A box of material 1 and density $\rho_1$ is surrounded by an ambient medium of material 2 and density $\rho_2<\rho_1$ in pressure equilibrium.
If the code does not properly reproduce the contact discontinuity, the pressure at the material interface becomes discontinuous. This creates an artificial surface tension \citep{Price2008} that, in turn, rounds the box's corners.
This creates an artificial surface tension \citep{Price2008} that rounds the box’s corners.
The total size of the computational domain in our simulation is $L=1$~\RE with periodic boundary conditions. The box ($-0.25 < x < 0.25$, $-0.25 < y < 0.25$ and $-0.25 < z < 0.25$, $\rho_1=20$ and internal energy $u_1=5$) is composed of iron and surrounded by a granite ambient medium with $\rho_2=10$ and internal energy $u_2=6.41092$ (all quantities are in code units). 
These initial conditions are then evolved with our SPH code with different SPH flavours for 17~h (in simulation time). The results are shown in Figure~\ref{fig:boxtest_iron_granite}.
In the simulation with classic SPH (i.e., without any modifications that improves the method's behaviour at interfaces) the box quickly transforms into a circle.
For the next simulation we use the geometric density average of the pressure forces (GDF) form of the SPH momentum equation proposed by \citet{Wadsley2017}.
This method reduces errors in the cases of strong density jumps and they found that in case of an ideal gas it substantially improves SPH's performance in the box test. 
When this method is applied to a non-ideal EOS like the Tillotson EOS GDF reduces the surface tension but only in combination with a correct density estimate at the interface (proposed in Section \ref{subsection:densitycorrection}).
Then the box remains stable over the entire simulation time (17~h) and the corners of the box are very well-resolved.

%%%%%%%%%%%%%%%%%%%%%%%%%%%%%%%%%%%%%%%%%%%%%%%%
% Fig: Box test for iron and granite 
%%%%%%%%%%%%%%%%%%%%%%%%%%%%%%%%%%%%%%%%%%%%%%%%

%%%%%%%%%%%%%%%%%%%%%%%%%%%%%%%%%%%%%%%%%%%%%%%%
% Issues with negative pressure.
%%%%%%%%%%%%%%%%%%%%%%%%%%%%%%%%%%%%%%%%%%%%%%%%
\section{Details on the interface correction}
\label{appendix:interf_correction}
The density correction requires a determination of the density ratio between the different materials for a given pressure and temperature. Generally this can only be done numerically by finding the root of $P \left( \rho, T \right) - P = 0$. Obtaining a unique solution requires a monotonically increasing pressure with increasing density in the region of interest. This is usually the case because
\begin{equation}
    \left.  \frac{\partial P}{\partial \rho} \right|_{T} > 0,
	\label{eqn:dpdrho}
\end{equation}
is a required condition for thermodynamical consistency of any EOS. 
There is a region in the expanded, cold states where the Tillotson EOS returns a negative pressure attempting to model tensile forces in a solid \citep{Melosh1989}. 
Since this is clearly unphysical for a fluid, and these negative values affect the numerical stability of SPH, the pressure is set to zero in these cases \citep{Reinhardt2017}.
To avoid complications with the root finder we allow for negative pressures in the EOS routine when inverting $P\left( \rho, T \right)$ and apply the "pressure cut-off" only when we calculate the particle's accelerations.
The interface correction is applied only when the obtained densities have a positive pressure for both materials.

\begin{figure}
	\centering
	\includegraphics{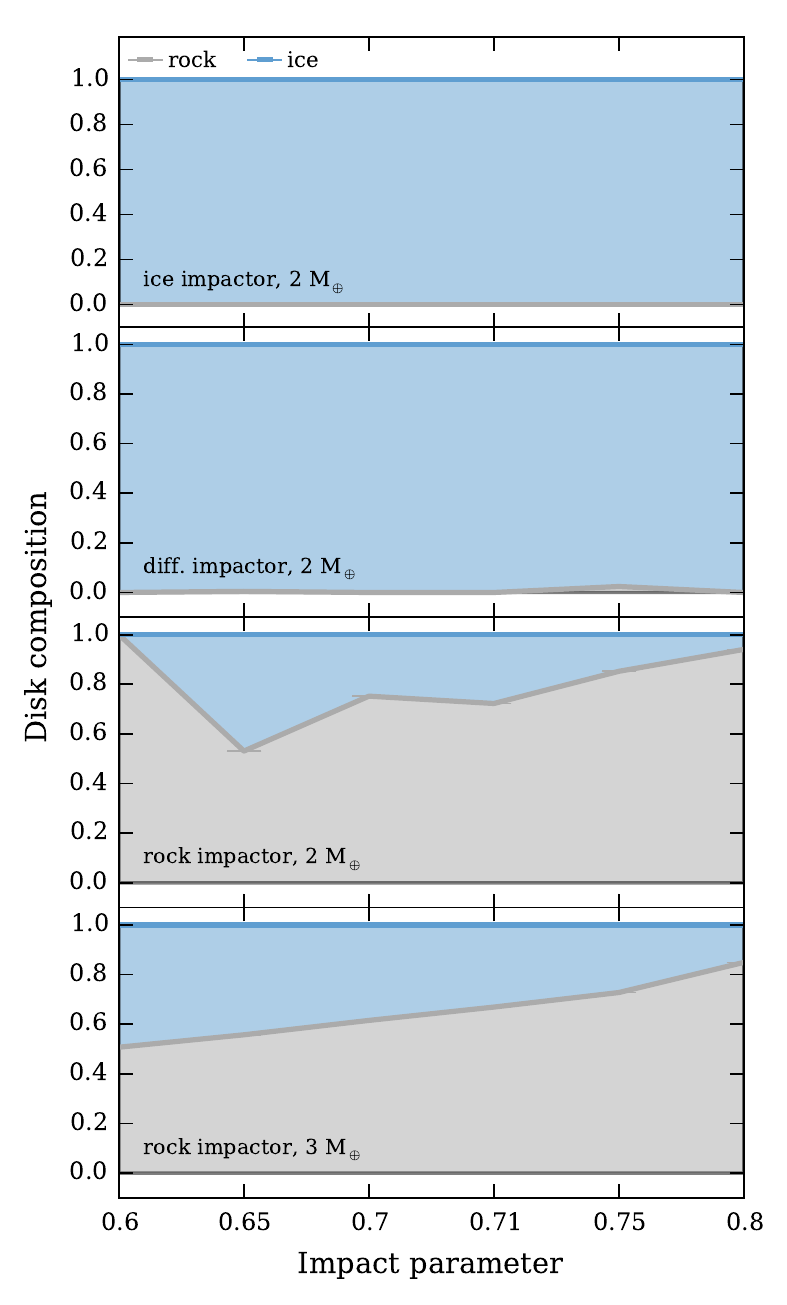}

	\caption{\textbf{The heavy-element composition of the proto-satellite disk.}
	Shown are the mass fractions of rock and ice in the proto-satellite disks presented in Figure~\ref{fig:uranus1_disk_composition}.
	In our simulations only a pure-rock impactor can produce disks that are substantially enriched in rock (for details see caption of  Figure~\ref{fig:uranus1_disk_composition}). For larger impact parameters a larger fraction of the  orbiting material originates from the impactor and therefore more rock is deposited in orbit.
	}
	\label{fig:uranus1_disk_composition_solid}
\end{figure}

%%%%%%%%%%%%%%%%%%%%%%%%%%%%%%%%%%%%%%%%%%%%%%%%
% Table of all simulations
%%%%%%%%%%%%%%%%%%%%%%%%%%%%%%%%%%%%%%%%%%%%%%%%
\include{table}

\end{appendix}
\clearpage
\bibliography{bib_uranus_neptune}{}
\bibliographystyle{mnras}

\end{document}

%% file: table.tex
\begin{table*}
    % MNRAS expects the caption before the table
	\caption{Table of all the simulations. The capital letters ABC in the ID number indicate the target's resolution: A) $10^5$ particles B) $10^6$ particles C) 5$\cdot 10^6$ particles. The capital letters UN stand for: U) Uranus and N) Neptune. The third and fourth characters stand for the impactor's mass: 1, 2 or 3~\ME and impactor's composition: i) for ice g) for granite d) differentiated, except for the hit-and-run collisions (HR) with same mass bodies. A run number follows this which usually refers to differing impact parameters in the order $0.1,0.2,0.3,0.4,0.5,0.6,0.65,0.7,0.71,0.75,0.8,0.9$ (12 impact parameters), unless otherwise specified. For HR collisions, we also indicate their rotation period a) P=0 hrs b) P=25 hours c) P=30 hours. To indicate the SPH flavors, we note for simulations run with: N) the density correction B) the Balsara switch I) the isentropic formalism W) the interface correction and P) the geometric density average of the pressure forces (see Appendix \ref{appendix:box_test}). \label{table:allsims}}
%	\begin{minipage}{150mm}
\begin{tabular}{ l | l | l | l | l }
%\begin{longtable}{ l | l | l | l | l }
%{\small
	ID & $b$ & $v_i$ [km/s] & SPH & CPU hrs\\ \hline
	AU1i01--12 & 0.1--0.9 & 20.16 & N & 7200 \\ 
	AU1g01--12 & 0.1--0.9 & 21.23 & N & 7220 \\
	AU1d01--12 & 0.1--0.9 & 20.25 & N & 7200 \\
	AU2i01--12 & 0.1--0.9 & 19.38 & N & 7200 \\
	AU2i13--24 & 0.1--0.9 & 19.38 & N,I,W,P,B & 7200 \\ 
	AU2g01--12 & 0.1--0.9 & 20.45 & N & 7220 \\ 
	AU2g13--24 & 0.1--0.9 & 20.45 & N,I,W,P,B & 7200  \\ 
	AU2d01--12 & 0.1--0.9 & 19.48 & N & 7200 \\ 
	AU2d13--24 & 0.1--0.9 & 19.48 & N,B & 7220 \\ 
	AU2d25--36 & 0.1--0.9 & 19.48 & N,P & 7220 \\ 
	AU2d37--48 & 0.1--0.9 & 19.48 & N,P,B & 7220 \\ 
	AU2d49--60 & 0.1--0.9 & 19.48 & N,I,W & 7220 \\ 
	AU2d61--72 & 0.1--0.9 & 19.48 & N,I,W,P & 7220 \\ 
	AU2d73--84 & 0.1--0.9 & 19.48 & N,I,W,P,B & 7220 \\ 
	AU3i01--12 & 0.1--0.9 & 19.33 & N & 7200 \\ 
	AU3g01--12 & 0.1--0.9 & 21.23 & N & 7220 \\ 
	AU3d01--12 & 0.1--0.9 & 19.51 & N & 7200 \\ \hline
	BU1i01--12 & 0.1--0.9 & 20.16 & N & 186624\\ 
	%BU1i13--24 & 0.1--0.9 & 20.16 & N,I,W,P,B & 7220 \\  
	BU1g01--12 & 0.1--0.9 & 21.23 & N & 196992\\ 
	%BU1g13--24 & 0.1--0.9 & 21.23 & N,I,W,P,B &  \\ 
	BU1d01--12 & 0.1--0.9 & 20.25 & N & 186624 \\ 
	BU2i01--12 & 0.1--0.9 & 19.38 & N & 186624 \\ 
	BU2i13--24 & 0.1--0.9 & 19.38 & N,I,W,P,B & 300672 \\ 
	BU2g01--12 & 0.1--0.9 & 20.45 & N & 196992 \\ 
	BU2g13--24 & 0.1--0.9 & 20.45 & N,I,W,P,B & 315187 \\ 
	BU2d01--12 & 0.1--0.9 & 19.48 & N & 186624 \\ 
	BU2d12--24 & 0.1--0.9 & 19.48 & N,B & 213408 \\ 
	BU2d25--36 & 0.1--0.9 & 19.48 & N,I,W,P & 300672 \\ 
	BU2d37--48 & 0.1--0.9 & 19.48 & N,I,W,P,B & 300672 \\ 
	BU3i01--12 & 0.1--0.9 & 19.33 & N & 186624  \\ 
	BU3i13--24 & 0.1--0.9 & 19.33 & N,I,W,P,B & 300672 \\ 
	BU3g01--12 & 0.1--0.9 & 21.23 & N & 196992 \\ 
	BU3g13--24 & 0.1--0.9 & 21.23 & N,I,W,P,B & 315187 \\ 
	BU3d01--12 & 0.1--0.9 & 19.51 & N & 186624 \\ 
	BU3d13--24 & 0.1--0.9 & 19.51 & N,I,W,P & 300672 \\ 
	BU3d25--36 & 0.1--0.9 & 19.51 & N,I,W,P,B & 300672 \\ \hline
	CU2i01--08 & 0.2,0.3,0.6--0.8 & 19.38 & N & 832000 \\ 
	CU2g01--08 & 0.2,0.3,0.6--0.8 & 20.45 & N & 919296 \\ 
	CU2d01--9 & 0.2,0.3,0.5,0.6--0.8 & 19.48 & N & 972000\\
	CU2d10 & 0.2 & 19.48 & N, BS & 108000  \\
	CU2d11--13 & 0.2,0.65,0.7 & 19.48 & N,I,W, & 324000 \\
	CU2d13--15 & 0.2,0.65,0.7 & 19.48 & N,I,W,P, & 324000\\
	CU2d16--18 & 0.2,0.65,0.7 & 19.48 & N,I,W,P,B & 324000\\ \hline \hline
	AN2i01--06 & 0.0--0.5 & 21.12 & N & 4200 \\	
	AN2g01--06 & 0.0--0.5 & 22.32 & N & 4210 \\
	AN2d01--06 & 0.0--0.5 & 21.22 & N & 4200 \\ 
	AN3d01--06 & 0.0--0.5 & 21.12 & N & 4200 \\ \hline
	CU2d01 & 0.2 & 21.22 & N & 108000  \\
	CU2d02 & 0.2 & 21.22 & N,I,W,P,B & 108000  \\
	\hline \hline
	AUHRa & 0.6 & 44.06 & N & 216 \\
	AUHRb & 0.6 & 44.06 & N & 216 \\
	AUHRc & 0.6 & 44.06 & N & 216 \\
	AUHRa & 0.7 & 44.06 & N & 216 \\
	AUHRc & 0.7 & 44.06 & N & 216 \\
%	}
%	\end{longtable}
\end{tabular}
%\end{minipage}
\end{table*}